\documentclass[conference]{IEEEtran}
\IEEEoverridecommandlockouts
\usepackage{cite}
\usepackage{soul}
\usepackage{amsmath,amssymb,amsfonts}
\usepackage[boxed]{algorithm}
\usepackage[noend]{algpseudocode}
\usepackage{paralist}
\usepackage{epsfig} 
\usepackage{graphicx}
\usepackage[export]{adjustbox}
\usepackage{cite}
\usepackage{subcaption}
\usepackage{listings,multicol}
\usepackage{lipsum}
\usepackage{setspace}
\usepackage{comment}
\usepackage{multirow}
\usepackage{tcolorbox}
\usepackage{color}
\usepackage{color, colortbl}
\usepackage{textcomp}
\usepackage{enumitem}
\usepackage{epstopdf}
\usepackage[font={small}]{caption}
\usepackage{wrapfig,lipsum,booktabs}

\definecolor{listinggray}{gray}{0.9}
\definecolor{lbcolor}{rgb}{0.9,0.9,0.9}
\newtheorem{Definition}{Definition}

\newtheorem{claim}{Claim}

\newcommand{\mc}[1]{\mathcal{#1}}
\newcommand{\mb}[1]{\mathbb{#1}}
\newcommand{\aawsos}{\emph{AAwSOS} }
\newcommand{\hns}{\emph{Hide-n-Seek} }
\newcommand{\twoNorm}[1]{\vert\vert{#1}\vert\vert}
\newcommand{\norm}[1]{\vert{#1}\vert}
\newcommand{\ceil}[1]{\lceil{#1}\rceil}

\newcommand{\lyapIT}[2]{V_{#1}(x(#2))}

\newcommand{\sigmaT}[1]{\sigma(#1)}

\def\BibTeX{{\rm B\kern-.05em{\sc i\kern-.025em b}\kern-.08em
    T\kern-.1667em\lower.7ex\hbox{E}\kern-.125emX}}
\begin{document}

\title{Concealing CAN Message Sequences to Prevent Schedule-based Bus-off Attacks
}

 \author{\IEEEauthorblockN{Sunandan Adhikary
  \IEEEauthorrefmark{1},
 Ipsita Koley
  \IEEEauthorrefmark{1},
  Arkaprava Sain\IEEEauthorrefmark{1},
  Soumyadeep das \IEEEauthorrefmark{3},
  Shuvam Saha \IEEEauthorrefmark{4},
 Soumyajit Dey
  \IEEEauthorrefmark{1}
 }
 \IEEEauthorblockA{
  \IEEEauthorrefmark{1}
 IIT Kharagpur, India
  ,\IEEEauthorrefmark{3}NIT Durgapur, India ,\IEEEauthorrefmark{4}Jadavpur University, India
 }
 }

\maketitle
\begin{abstract}
This work focuses on eliminating timing-side channels in real-time safety-critical cyber-physical network protocols like Controller Area Networks (CAN). Automotive Electronic Control Units (ECUs) implement predictable scheduling decisions based on task level response time estimation. Such levels of determinism exposes timing information about task executions and therefore corresponding message transmissions via the network buses (that connect the ECUs and actuators). With proper analysis, such timing side channels can be utilized to launch several schedule-based attacks that can lead to eventual denial-of-service or man-in-the-middle-type attacks. To eliminate this determinism, we propose a novel schedule obfuscation strategy by \emph{skipping} certain control task executions and related data  transmissions along with \emph{random shifting} of the victim task instance. While doing this, our strategy contemplates the performance of the control task as well by bounding the number of control execution skips. We analytically demonstrate how the \emph{attack success probability (ASP)} is reduced under this proposed \emph{attack-aware skipping and randomization}. We also demonstrate the efficacy and real-time applicability of our \emph{attack-aware schedule obfuscation strategy} \hns by applying it to synthesized automotive task sets in a real-time Hardware-in-loop (HIL) setup. 
\end{abstract}


\section{Introduction}
\label{sec:intro}
The development cycle of real-time cyber-physical systems (CPSs), like automotive, generally considers safety and timeliness as their first-class design criteria. This along with the limited availability of  computation/communication resources often lead to adoption of predictable static/dynamic task scheduling algorithms. 
  As a result, we have deterministic task execution schedules (e.g. Autosar classic platforms) involving  multiple control tasks for several CPS closed loops running in shared real-time embedded platforms. Achieving such predictable timing and functionality while  considering the \emph{worst case} behaviour is also desired by the system designers in order to ensure the safety of the closed-loop in the presence of delay, jitters and interferences in shared processors and communication mediums. It may be noted that platform-level deterministic task execution leads to regular patterns of messages in the associated  communication medium  that the processing units use for sensing, actuation as well as inter-ECU data sharing. This  predictability in task and message schedules open up timing side-channels\cite{hounsinou2021vulnerability} in processor and communication channels.

\par
In modern automotive systems, modular design of the application layers, runtime environments, and communication stack designed following the AUTOSAR standards \cite{autosar_13}, highly restricts on-board vulnerabilities from exposure. Therefore, the lightweight in-vehicle communication protocols have become ideal targets of a \emph{schedule-based attacker}.
The most preferred target for such \emph{man-in-the-middle} type attacks is the Control Area Network (CAN) \cite{koscher2010experimental}, the de-facto intra-vehicular communication protocol that transmits most of the \emph{real-time safety-critical messages} among the 
electronic control units (ECUs). Typically, tasks in vehicle ECUs follow a static schedule whenever arrival of new tasks does not occur.
The related messages produced by the control tasks are priority-wise communicated via CAN bus following a static schedule. 
Since CAN protocol lacks any authentication and confidentiality scheme, it becomes an ideal target for an attacker. The deterministic timing behaviour of the safety-critical CAN messages is utilized to launch schedule-based attacks that can disconnect an ECU temporarily by sending it to \emph{bus-off} mode (known as \emph{bus-off attack})~\cite{cho2016error,hounsinou2021vulnerability}. In a bus-off attack, the attacker is essentially a piece of code implemented on a compromised ECU. It analyzes the CAN traffic, identifies the periodicity and arrival times of a targeted message, fabricates an attack message with the same identifier as the targeted one. The attacker injects the attack message into CAN synchronously with the target message at the same time. Repeated synchronous transmissions cause the transmission error count of the victim message's source ECU to increase and gradually send the victim ECU to the bus-off mode. The bus-off attack is one of the most lethal attacks on CAN protocol proposed in the last decade because the attacker remains stealthy during the attack process as it uses the same identifier and timing information as the victim message. With the victim ECU in bus-off mode, the attacker can now transmit false data messages mimicing the victim and causing further damage to the system.

\par A significant number of solutions have been proposed to counter bus-off attacks. 
One line of work uses physical properties, like clock skew, and voltage profile, of the ECUs to identify the source of all the messages beforehand \cite{cho2017viden, cho2016fingerprinting,kulandaivel2019canvas}. At run-time, if a message exhibits different values of such properties, it is flagged as coming from a compromised ECU. The primary limitation of these approaches is that they take significant time to detect bus-off attack attempts. On the other hand, once the synchronization with the victim message is achieved, the entire bus-off attack takes very less time to accomplish. Another line of intrusion detection systems proposed to encounter bus-off attacks is the use of trusted modules to provide authentication to the transmitted messages \cite{ying2019tacan,jo2019mauth,serag2023zbcan}. However, these solutions incur additional installation costs. Moreover, if the trusted module is compromised, the entire intrusion detection system fails. In this work, we propose a lightweight attack-aware schedule randomization strategy \hns to obfuscate ($hide$) the CAN schedule from timing leakages as well as detect ($seek$) the bus-off attack attempts. 
\emph{Schedule randomization} is largely suggested as a helpful strategy to defend against processor-level timing attacks~\cite{kruger2018vulnerability, vreman2019minimizing}. Randomized schedules can obfuscate the timing information about the schedule and theoretically promise higher entropy. 
However, it has also been shown by some researchers that the increment in the entropy of schedule randomization cannot ensure reduced \emph{attack success probability (ASP)}. Rather randomizing without any insight into the attacker's strategy might increase the penetrability of the  system~\cite{nasri2019pitfalls}. 

In this work, we propose an attack-aware schedule obfuscation strategy (\aawsos)  that instruments an ECU schedule that hides the timing information of safety-critical messages, transmitted through the CAN bus, from an attacker who is attempting a schedule-based attack (SBA), like bus-off attack. 
The proposed \aawsos module in each ECU $e$ will first monitor the traffic for a certain number of CAN hyper-periods. Next, it will find out which instances of the safety-critical control tasks, transmitted from $e$, are most vulnerable to bus-off attack w.r.t. this analysis window. \aawsos then either \textbf{(a)} skips those vulnerable instances of the safety-critical control tasks, or \textbf{(b)} skips one or more instances of some higher priority control tasks in $e$. 
In general, control theorists design mathematical control laws while keeping environmental and platform delays in mind. This generally implies the control loop will work with the desired performance guarantee up to a bounded number of delay-induced control task execution skips  \cite{ghosh2017structured}. 
In this work, we propose to intentionally inject such skips with security as a goal in mind. Switching between many such control skipping sequences makes information leakage about the execution decision of task instances
less probable as the positions of job skips for the same task keep changing. Consequently, the faulty schedule analysis exposes the attacker. However, skipping an arbitrary number of control executions can degrade the performance of the safety-critical systems. To ensure performance, \aawsos cannot skip consecutive control tasks beyond a certain limit. In those situations, \aawsos \textbf{(c)} will change the ordering of the same priority tasks in $e$ such that none misses the respective deadline to fool the attacker's schedule analysis. We show that schedule obfuscation techniques (a), (b), and (c) reduce the ASP much less than the blind schedule randomization techniques. 
The primary contributions of this work are summarized below:\\
    \textbf{(1)} We propose an attack-aware schedule obfuscation strategy (\aawsos) \hns to {\em hide} the timing information of safety-critical control tasks transmitted through the CAN bus from an attacker attempting to launch a bus-off attack. The proposed strategy also detects possible bus-off attempts while obfuscating the schedules.
    It also ensures that a desired performance guarantee is always ensured and none of the tasks misses their deadlines.\\
    \textbf{(2)} We mathematically establish that \aawsos will reduce the attack success probability (ASP) by a significant amount when compared to static schedules or attack-unaware schedule randomization techniques.\\
    \textbf{(3)} We evaluate the efficacy of the proposed \hns strategy to combat and detect CAN bus-off attacks on a hardware-in-loop (HIL) setup using benchmark CAN traffic data.
\section{Background}
\label{sec:bg}
\subsection{Intra-vehicular Network}
\label{subsec:intraVehNtwk}
\begin{figure}[!ht]
    \centering
    \includegraphics[width=\columnwidth]{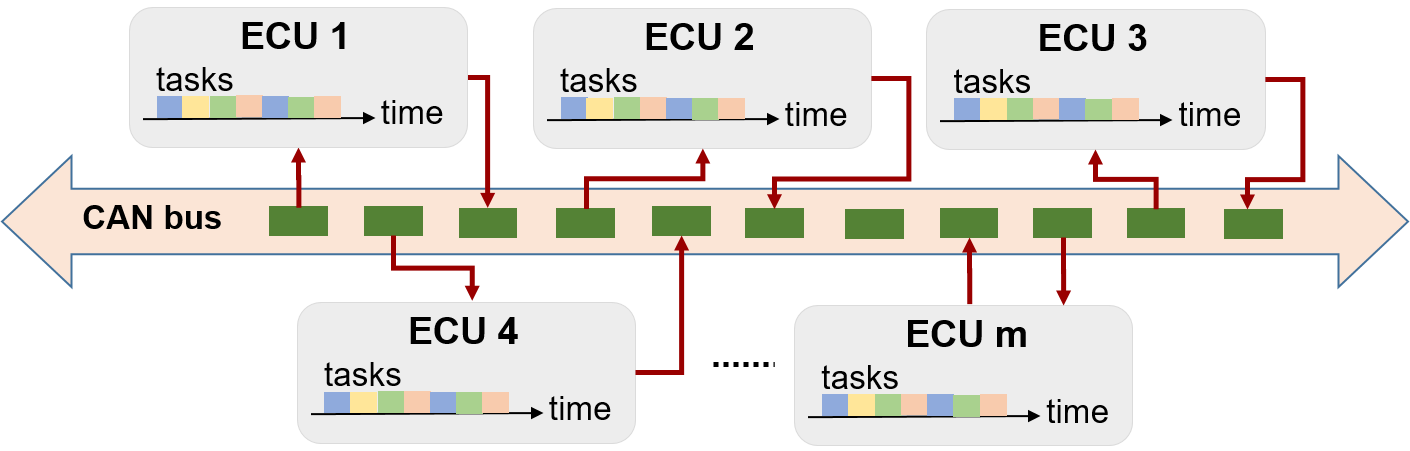}
    \caption{Intra-vehicular Network}
    \label{fig:intraVehNtwk}
\end{figure}
We consider a typical automotive closed-loop setup with a CAN network connecting the plants and different ECUs. Such a network is demonstrated in Fig.~\ref{fig:intraVehNtwk}. We consider there are multiple 
ECUs connected to the CAN bus.
Each ECU has multiple cores and a number of tasks (including safety-critical ones) are running on them following certain static schedules. We assume one of the cores of ECUs is always available to run the proposed \hns algorithm. The execution of this algorithm is triggered by an interrupt that is raised whenever a message is transmitted in the CAN bus enabling it to observe the CAN traffic. This is considered practical as most of the modern ECUs (eg. Infineon Tricore-397, Raptor GCM-196 etc.) have multi-core implementations.
One or more tasks running on ECUs transmit and receive messages to and from the CAN bus. On arrival, such a task instance is executed in the ECU following a static schedule (since the set of tasks on an ECU is fixed), and the data produced by the task is attempted for transmission via CAN as a CAN packet/frame. So the relative order of the ECU tasks and the messages transmitted by them in CAN remains intact. 

CAN packets have no source and destination addresses and are communicated in \emph{broadcast} mode. Each data packet has a unique ID that signifies its transmission priority in the bus. Contention to the CAN bus by multiple messages at the same time is resolved through the \emph{CAN arbitration} process. Technically CAN works as a \emph{wired-AND gate}. An ECU sending a message with a higher ID value will see a \emph{dominant bit i.e.} $0$ in the bus if another ECU sends a message with a lower ID value at the same time, and thereby, retrieves itself from transmitting. Thus, \emph{a message with a lower ID value is considered to be of higher priority over a message with a higher ID value}. 
 In this work, we focus on the \emph{safety-critical control tasks} that are executed in the ECUs and the control messages transmitted by them via the CAN bus. To understand their importance, we provide a background on typical 
 automotive control tasks, their modeling, and stability/performance requirements.
\begin{figure}[!ht]
    \centering  
    \includegraphics[width=0.95\columnwidth,clip]{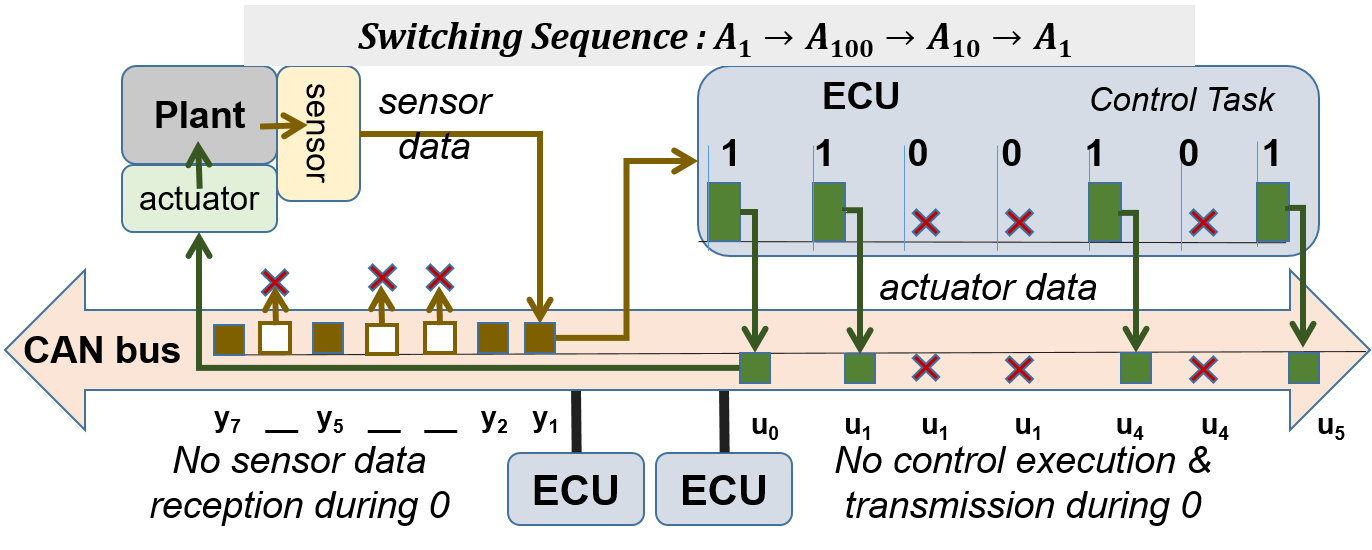}
    \caption{An embedded CPS following a CSS \textbf{1101001}}
    \label{fig:sysundercss}
    \vspace{-4mm}
\end{figure}
\subsection{Closed-loop Safety Critical CPS Model}
\label{subsec:cpsModel}
Closed-loop safety-critical automotive CPSs are implemented across multiple ECUs in modern vehicles. Such CPS closed-loops are in charge of different safety-critical real-time operations e.g. anti-lock braking system (ABS), traction control, electronic stability control (ESC), adaptive cruise control (ACC), etc. The controller and the actuator units are implemented in different ECUs that are connected via the CAN bus. The state-space equations for a discrete-time linear-time-invariant (LTI) closed-loop system in Eq.~\ref{eq:nodrop} capture the progression of this plant-controller closed-loop. The controller units rely on the sensed measurement data $y$ from the plant, which is communicated through the CAN. Using this an estimator task first estimates the states of the plants. We denote $x$ as the plant's state vector, and $\hat{x}$ as the estimated state vector by the estimator. Based on the estimated plant state, a control task calculates the required control input $u$ in order to transmit it to the actuator for actuation and control of the plant.
\begin{align}
\label{eq:nodrop}
\nonumber
    &x[k + 1] = A x [k] + B u[k],\ \  y[k+1]= C x[k+1]\\
\nonumber
    &\hat{x}[k + 1] = (A-LC) \hat{x} [k] + B u[k] + L y[k]\\ &u[k+1] = K \hat{x} [k+1]    
\end{align}
Here, $A, B, C$ are characteristic matrices. $K$ and $L$ are the controller gain and gain of the Kalman estimator. Indexing the system variables with $k$ simply denotes the sampling iteration of the discrete LTI system. 
Considering the augmented state vector of the plant-controller closed-loop as $X= [x^T,\hat{x}^T]^T$, and by replacing $u$ with $K\hat{x}$, $y$ with $Cx$ the overall closed-loop system evolves as follows: 
%
\begin{align}
    X[k+1]= A_{1}X[k],\ 
    {\footnotesize A_{1}=\left[{ \begin{array}{cc}A&  B K\\L C& A-L C+B K\end{array}}\right]}
    \label{eq:A1}
\end{align}
We now, introduce the idea of \emph{skipping a control execution} and observe how the aforementioned closed-loop evolves in the presence of such a skip. Skipping a control execution at $(k+1)$-th time instance means the progression of the estimator-controller unit is stalled at that iteration, i.e. $\hat{x}[k+1]=\hat{x}[k]$. Therefore, no sensor data with the output measurement of $k$-th sampling iteration is transmitted through CAN, and consequently, no new control messages are computed and transmitted at $(k+1)$-th sampling iteration as well. Therefore, at $(k+1)$-th iteration the plant evolves using the last-computed control input i.e. $u[k+1] = u[k]$. In presence of a control execution skip at $(k+1)$-th iteration, the closed loop system in Eq.~\ref{eq:nodrop} evolves as follows. 
\begin{align}
\label{eq:drop}
    &x[k + 1] = A x [k] + B u[k],\ \  
    \\
\nonumber
    &\hat{x}[k + 1] = I \hat{x} [k] + O u[k] + O y[k],\ \  u[k+1]= K \hat{x} [k+1]    
\end{align}
$I$ and $O$ are identity matrices and zero matrices with the same dimensions as $A$, $B$ respectively. The augmented closed-loop system under control execution skip progresses like:
\begin{align}
    X[k+1]= A_{0}~X[k],\ {\footnotesize A_{0}=\left[{ \begin{array}{cc}A&  BK\\O& I \end{array}}\right]}
    \label{eq:A0}
\end{align}
\noindent Due to the occasional control skips, we thus have a switched LTI system where we switch between the different combinations of $\{A_1,A_0\}$. For example, if we execute the control task at $k$-th iteration and skip the control executions at $(k+1)$-th and $(k+2)$-th iterations, the augmented closed-loop system state at $(k+3)$-th iteration becomes $X[k+3]=A_0A_0A_1X[k]$. This can be represented as a subsystem with state transition matrix $(A_0)^2A_1$. Switching between such subsystems gives a control execution skip sequence, formally defined below.
\begin{Definition}[\bf Control Skipping Sequence]
An $l (\in \mb{N})$ length control skipping sequence (CSS) for  a given control loop is a sequence $\rho\in \{0, 1 \}^l$ such that 
$1$ denotes a control execution when the augmented state of the plant and controller $X$ progresses following Eq.~\ref{eq:A1}, and $0$ denotes a skipped execution when $X$ progresses following Eq.~\ref{eq:A0}.
\label{def:ctrlskippat}\hfill$\Box$
\end{Definition} 
A control skipping sub-sequence of an $l$-length control skipping sequence $\rho$ is an $i$-length sub-string of $\rho$ having the form $10^{i-1}$ such that $\sum_q{i}=l$, where $q$ denotes the number of such sub-sequences in a sequence. A sub-sequence signifies that the plant is actuated with the last control update in consecutive $(i-1)$ iterations. As an example, the control execution sequence $\rho= 1100101$ is generated by 4 times switching between 3 \emph{control skipping sub-sequences} i.e. $1$ to $100$, $100$ to $10$ and back to $1$ again, i.e. {\small$X[7] = A_{1}A_{10}A_{100}A_{1}X[0]= (A_{0})^{0}A_{1}(A_{0})^{1}A_{1}(A_{0})^{2}A_{1}(A_{0})^{0}A_{1}X[0]$}. Fig.~\ref{fig:sysundercss} demonstrates the transmission of sensor data and control signals through CAN for an automotive CPS during this control skip sequence. 
\noindent\par $\bullet$ \textbf{\em Performance-aware Switching between Control Skipping Sub-sequences:}
In order to measure the effect of skipping control executions on the performance of the closed-loop system, we represent the augmented closed-loop system as a switched linear system like below.
\begin{align}
\label{eq:linearSwitchedSys}
X[k+1] = A_{\sigma[k]}(X[k])
\end{align} 
Here, the switching signal $ \sigma[k]=i$ signifies that at $k$-th sampling instance, the system follows the dynamics of the control skipping sub-sequence $10^{i-1}$, which we denote as the $i$-th subsequence. While following an $i$-th subsequence, a closed-loop skips control execution in consecutive $(i-1)$ iterations and reuses the last computed control update in all these iterations. We use Global Uniform Exponential Stability (GUES) as the performance-driven stability metric to ensure that a CSS maintains a given settling time requirement.
The equilibrium at $X = 0$ of a discrete LTI system attains GUES under a certain switching signal $\sigma[k]$ if, for initial conditions $X[k_0]$, there exist constants $M > 0, \ 0 < \gamma < 1$ such that {\small$||X[k]|| \leq M \gamma^{(k-k_0)}~||X[k_0]||, \ \forall k \geq k_0$} where $||.||$ is the vector norm and $\gamma$ is the GUES decay rate. To ensure that the switched closed-loop system attains certain GUES decay requirement, we {\em arbitrarily} switch between only those sub-sequences for which, there exists a {\em Common Lyapunov Function} (CLF). For \emph{arbitrarily switching between such control skipping sub-sequences} the following claim must hold.
\begin{claim}\label{thm:clf}
A switched system like Eq.~\ref{eq:linearSwitchedSys} can \emph{arbitrarily} switch between a $j$-th subsequence and an $i$-th subsequence if and only if a common Lyapunov function (CLF) $V$ exists for all such $j$-th and $i$-th subsequences. The CLF should satisfy the following CLF criteria in order to maintain a desired GUES decay margin of $\gamma$ while switching among them arbitrarily.
\begin{align}
\label{eqn:lyapKinf}
    &\kappa_{1}(\twoNorm{X(k)})\leq \lyapIT{i}{k} \leq \kappa_{2}(\twoNorm{X(k)})\\
\label{eqn:lyapDeriv}
    &\Delta V_{i}(X(k)) \leq \alpha_{i}\ \lyapIT{i}{k} \ 
 \text{and for some}\ \alpha_{i}\neq 0\\
\label{eqn:lyapMu}
    &\lyapIT{i}{k}\leq \mu_{i}\ \lyapIT{j}{k^{-}}\ \forall j,\text{s.t. }i\neq j,\ \mu_{i}=1\\\nonumber 
    &\kappa_{1},\kappa_{2}\in \text{ class }\mathcal{K}_{\infty},\ k^{-}< k, \sigmaT{k} =i,\ \sigma(k^{-})=j
    ~~~~~~~~\Box
\end{align}
\end{claim}
The Lyapunov functions for $i$-th and $j$-th subsequences($V_i,V_j$ respectively), are derived by solving the LMIs for a given desired GUES decay rate $\gamma$ as mentioned in the above claim. Among such sub-sequences with CLF, the one with maximum number of consecutive control execution skips gives us the upper bound or \emph{skip limit}. For example, consider for a closed-loop, a stable CSS given by $(1+10+100)^+$ (in regular language notation). The desired GUES decay is generated by arbitrarily switching between the subsequences, $1$, $10$, and $100$. In this example, $2$ is the \emph{skip limit} for the closed-loop as maximum two consecutive skips are allowed in the participating sub-sequences. While obfuscating CAN schedules by skipping  control tasks, the skip limit helps in preserving control performance of the related closed-loop. 

\subsection{Bus-off Attack on CAN: A Schedule-based Attack (SBA)}
\label{subsec:busoff}
\begin{figure}[!hb]
    \centering
    \includegraphics[width=0.95\columnwidth]{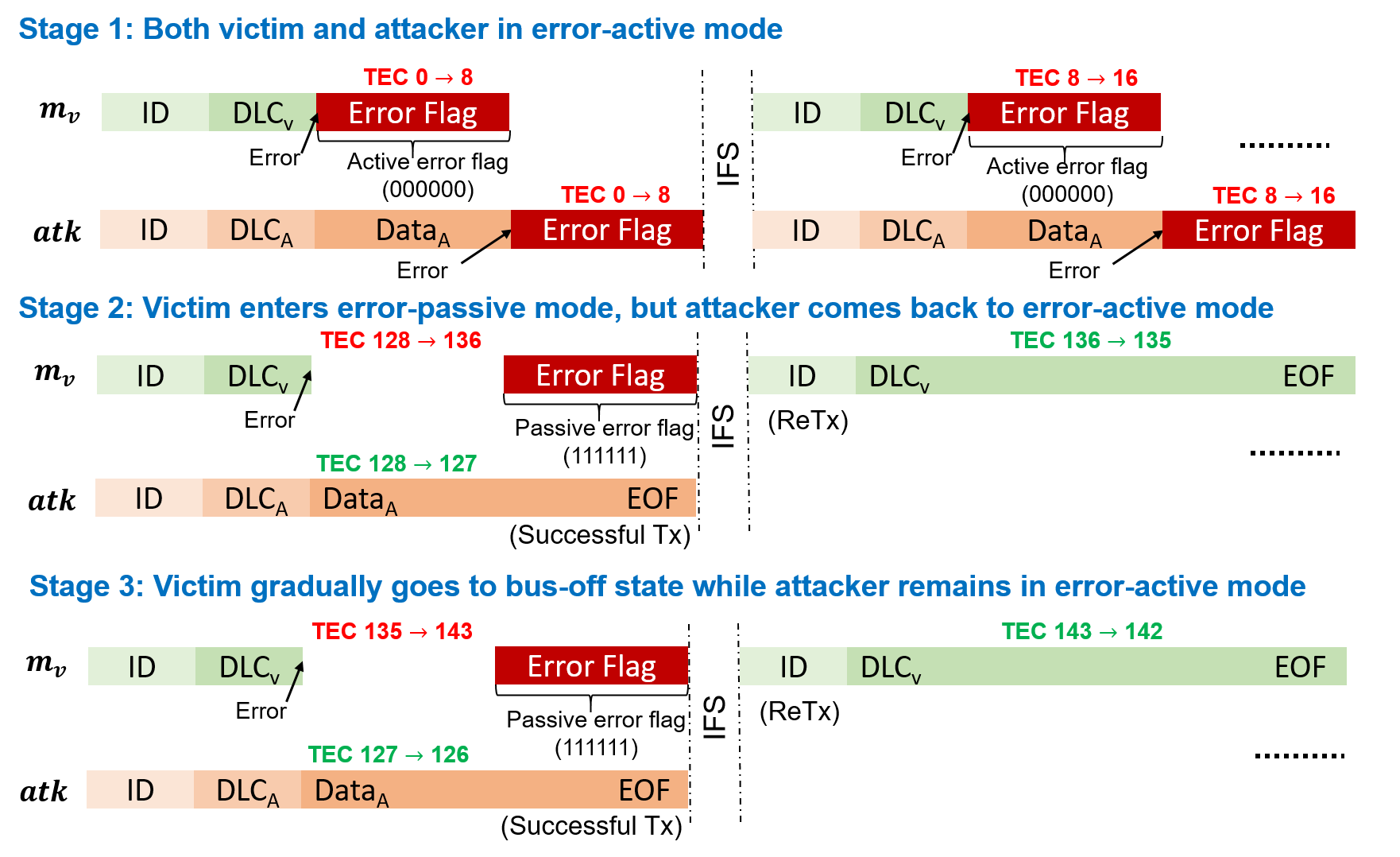}
    \caption{\centering{Three stages of bus-off attack\cite{cho2016error} (
    $m_v$: victim message, $atk$: attack message, $EOF$: end of frame, $IFS$: inter-frame space)}}
    \label{fig:busOff}
\end{figure}
In this work, we consider an attacker that attempts to launch a bus-off attack~\cite{cho2016error} on a victim message ID in the CAN bus. We consider the control messages, originating from the control tasks as the victims in our attack model. By targeting a control message, the attacker intends {\em to launch a bus-off attack on the victim ECU} i.e. the source of this control message ID (where the corresponding control task runs). 
The bus-off attack exploits the error-handling strategy of CAN protocol which states that an ECU (i) remains in {\em error active} mode if the transmission error count (TEC) or reception error count (REC) is less than $128$, (ii) enters an {\em error passive} mode when TEC or REC exceeds $127$, but remains below $256$, and (iii) goes to the {\em bus-off state} i.e. gets disconnected from the bus temporarily when  TEC or REC exceeds $255$ to notify some significant disruption in transmission. Before launching the bus-off attack, the attacker, which is fundamentally a piece of code implemented on a compromised ECU, targets a safety-critical periodic control message $m_v$ and monitors the CAN bus traffic to compute the timing parameters (periodicity, arrival offsets, etc.) of $m_v$. It fabricates attack messages $atk$ with the same ID as $m_v$ and a value in the data field less than the one in $m_v$. Before launching the bus-off attack, it is assumed that both the victim node i.e. the ECU transmitting $m_v$, and the attacker node, i.e. the compromised/attacker ECU are in error active mode.

\par The $3$ stages of the bus-off attack are demonstrated in Fig.~\ref{fig:busOff}. At the first stage, the attacker injects $atk$ in the CAN bus {\em synchronously} with $m_v$ i.e. exactly at the same time. Since both $atk$ and $m_v$ have the same ID, both win the arbitration process at the ID field. However, during the transmission of the data field, the transmitting ECU of $m_v$ senses the error first, and broadcasts an active error flag (containing $000000$). As CAN works as a wired-AND gate, attacker ECU senses transmission error as well. This leads to an increase in TEC by $8$ for both the victim and attacker ECUs. Due to the failed transmissions, both victim and attacker ECUs attempt retransmissions of their respective messages at the same time and the same transmission error occurs. After $16$ consecutive retransmission attempts, the TEC of both attacker and victim node reach $128$ i.e. both the nodes enter error-passive mode. Synchronization of $m_v$ and $atk$ in error-passive mode causes the victim ECU to send a passive error flag (containing $111111$) with a further increase in TEC by $8$. As this generates no error in the attacker node, the attack message is transmitted successfully by reducing its TEC by $1$. As an effect, the victim node's subsequent retransmission attempt becomes successful without any obstacle from the attacker and reduces its TEC by $1$. In the second stage of the attack, while the attacker is in error passive mode, it again synchronizes with the victim during the next transmission cycle. The attacker's TEC keeps on reducing by $1$, and the victim's TEC increases by $8-1=7$. At this third stage of the attack, after subsequent synchronous transmissions in error passive mode, the attacker comes back to the error active mode but the victim eventually goes into the bus-off state. In the next sections, we present our novel schedule obfuscation strategy \aawsos utilizing control execution skips. Our \aawsos can \emph{hide} the CAN bus schedule of a victim controller message and prevent the bus-off attack attempt from becoming successful while \emph{seeking} to detect the attacker.

\section{Proposed Methodology}
\label{sec:method}
Before we explain our attack-aware schedule obfuscation policies, we need to understand how a bus-off attacker looks for the {\em oppertunest} moments to launch a bus-off attack. This serves as a motivation behind the innovation of various schedule obfuscation policies in an attack-aware way.
\subsection{Schedule Analysis To Launch Bus-off Attack}
\label{subsec: schedAnalysis}
  \begin{figure*}[!ht]
    \centering
\includegraphics[width=1.9\columnwidth]{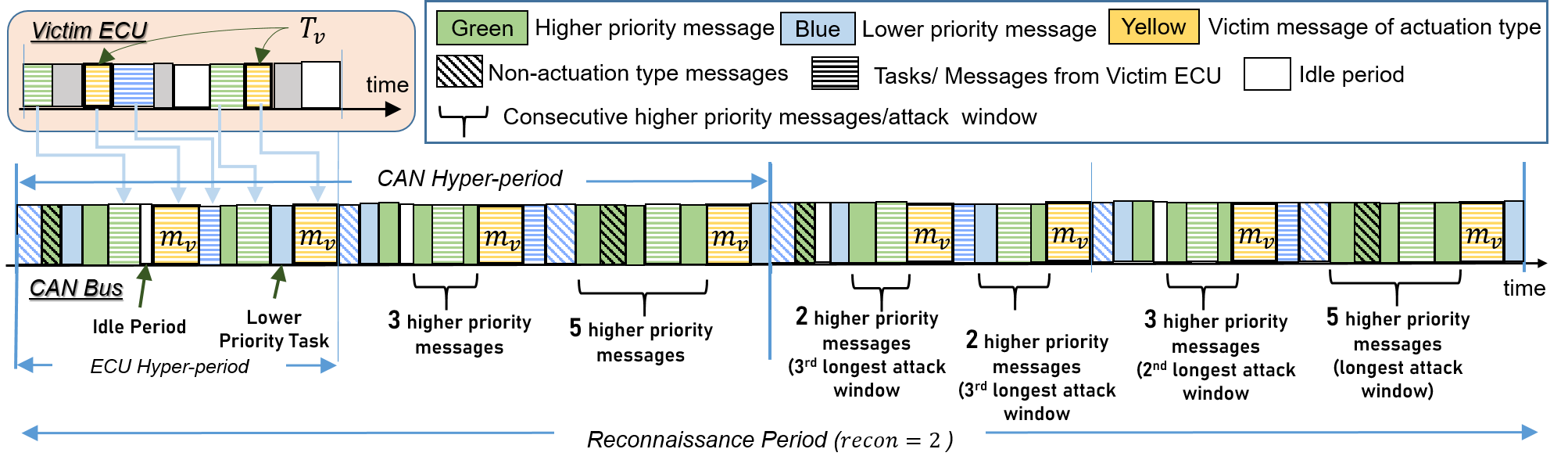}
    \caption{CAN schedule analysis for launching bus-off [the green/blue `unpatterned' messages signify control tasks in other than victim ECUs]}
    \label{fig:schedAnalysis}
\end{figure*}
\begin{figure*}[!ht]
    \centering
\includegraphics[width=1.9\columnwidth]{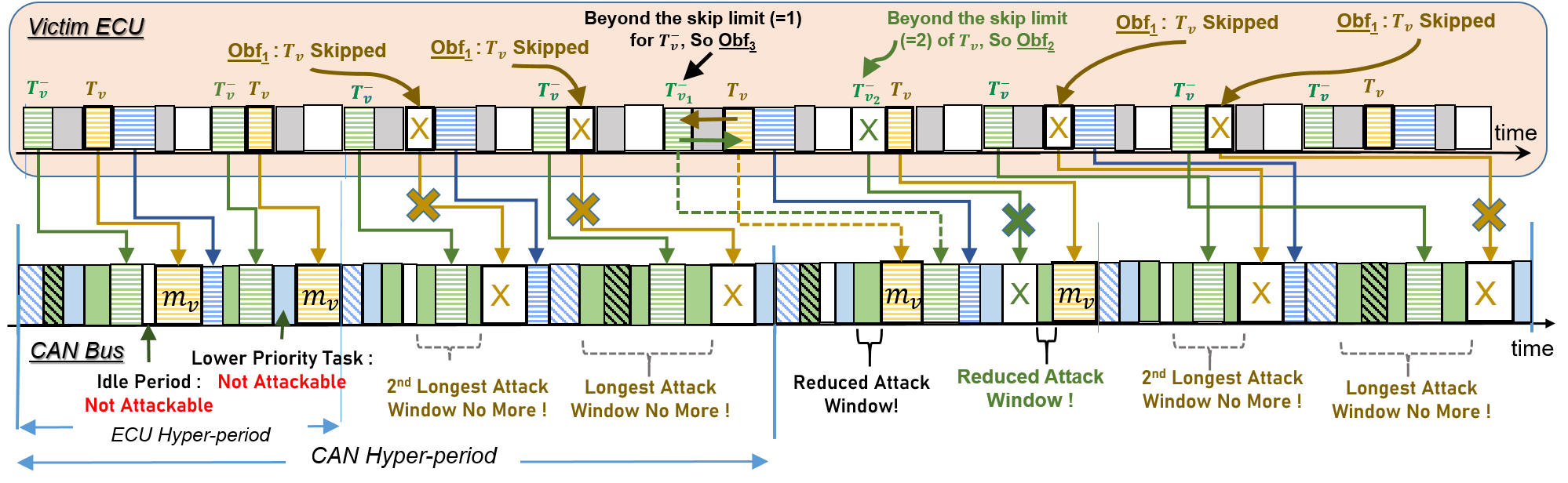}
    \caption{CAN schedule analysis for launching bus-off Under \hns [legends are same as Fig.~\ref{fig:schedAnalysis}]}
    \vspace{-4mm}
    \label{fig:obfschedAnalysis}
\end{figure*}
The success of the bus-off attack heavily relies on how precisely the attack message $atk$ synchronizes with the targeted instance of victim message $m_v$ (see Fig.~\ref{fig:busOff}). The attacker analyzes the CAN schedule to find out targeting which instance of $m_v$ results in the highest probability of synchronization, and thereby, bus-off attack \cite{hounsinou2021vulnerability}. As shown in Fig.~\ref{fig:schedAnalysis}, there can be $3$ types of message-sequences in CAN bus preceding each instance of $m_v$: i) one or more higher-priority messages, i.e. whose IDs are lower than $m_v$'s (e.g., $3$-rd and $4$-th instances of $m_v$ in the $1$-st hyper-period of CAN bus in Fig.~\ref{fig:schedAnalysis}); ii) one or more lower-priority messages, i.e. whose IDs are higher than $m_v$'s (e.g., the $2$-nd instance of $m_v$ in the $1$-st hyper-period of CAN bus in Fig.~\ref{fig:schedAnalysis}), and iii) idle time slot (e.g., the $1$-st instance of $m_v$ in the $1$-st hyper-period of CAN bus in Fig.~\ref{fig:schedAnalysis}). In the case of (iii), the attacker has to inject $atk$ at exactly the same time as $m_v$ to initiate the bus-off attack. Perfect synchronization of $atk$ with $m_v$ will be extremely difficult due to bus jitters. Therefore, the attack success probability (ASP) is nearly negligible in this scenario. In the case of (ii), the attack initiation window is, at most, the transmission time of the immediately preceding lower-priority message. Note that the attacker must precisely inject $atk$ after the lower-priority message wins the bus arbitration, i.e. after the transmission of the ID field of the lower-priority message. Otherwise, $atk$ will win the arbitration as it has higher priority. This makes the $atk$ face the CAN arbitration before $m_v$ and hence loses the intended synchronization with $m_v$. In the case of (i), the attacker can initiate the bus-off attack anytime during the transmission of one or more consecutive higher-priority messages preceding $m_v$. This will force $atk$ to synchronize with $m_v$ after the transmission of the last higher-priority message preceding $m_v$. Therefore, the ASP, in this case, is much higher than (ii) and (iii). 

\par In this schedule-based attack (SBA) setup, we consider the following two assumptions regarding the attack model. $First$, the attacker targets a messages $m_v$ corresponding to a safety-critical control task $T_v$. $Second$, it will always target those instances of an $m_v$ that immediately follow one or more higher-priority messages in the bus (i.e. type (i) message sequence mentioned in the previous paragraph). Therefore, we consider this sequence of high-priority tasks coming just before the instances of a victim message $m_v$ in the CAN bus as \emph{attack window}. 
The ASP corresponding to each instance of $m_v$ changes proportionally with the attack window length. The attacker observes the CAN schedule for a number of hyper-periods, say $recon$, to identify which instance of the victim message can be targeted for high ASP. We call this duration as \emph{reconnaissance period}. The attacker computes the average attack window length for each instance of $m_v$ over $recon$ CAN hyper-periods. The ASP will be highest if the attacker targets the particular instance of $m_v$ where the attack window is the longest. For example, in Fig.~\ref{fig:schedAnalysis}, the average numbers of higher-priority messages coming just before the $4$ instances of $m_v$ over $recon = 2$ hyper-periods in CAN bus are $\lceil(0+1)/2\rceil=1$, $\lceil(0+2)/2\rceil=1$, $\lceil(3+3)/2\rceil=3$, and $\lceil(5+5)/2\rceil=5$ respectively. Therefore, targeting the $4$-th instance of $m_v$ in the next reconnaissance period will lead to maximum ASP for an attacker. 
	
\subsection{Attack-aware Schedule Obfuscation Strategies}
\label{subsec:obfRules}
The primary motivation of this work is to devise a lightweight strategy to break the attacker's schedule analysis for finding the target instance of victim message $m_v$ that exhibits high ASP. In this regard, we propose $3$ attack-aware schedule obfuscation strategies (\aawsos). Based on the analysis of the CAN traffic during the current reconnaissance period, the \hns framework implemented in each ECU will instrument a task execution schedule for the corresponding ECU during the subsequent reconnaissance period. The objective of the \hns algorithm is to obfuscate the future control task schedules in an ECU to introduce a maximum reduction in ASP of the resulting global CAN schedule.  In Fig.~\ref{fig:schedAnalysis}, we provide a snapshot of the CAN bus in the current reconnaissance period whose length is taken as $2$ consecutive hyper-periods. By analyzing this CAN message schedule in the current reconnaissance period, \hns obfuscates the existing control task execution schedule that is to be followed during the next reconnaissance period as demonstrated in Fig.~\ref{fig:obfschedAnalysis}. This obfuscation decision is not a static one, as the CAN traffic changes due to $(i)$ the local obfuscation decisions taken by the other ECUs connected to the bus, $(ii)$ timing interference caused by the arrivals of `non-actuation' and aperiodic messages, and $(iii)$ obviously, due to the undesired delays and jitters. This demands a cyclical execution of our \hns algorithm for every reconnaissance period. The idea is to observe the current CAN traffic and instrument the next ECU schedule for every sliding window of the length same as the reconnaissance period. While doing so, the \hns algorithm observes the bus and flags the message instances corresponding to a scheduled skip in task execution and thereby detects a bus-off attack attempt.

\par \noindent$\bullet$ \textbf{Obfuscation Strategy 1 ($Obf_1$)} - \emph{Skipping the vicim task instance}: As discussed in Sec.~\ref{subsec:intraVehNtwk}, the relative orders of the tasks executed in the ECU's schedule and their corresponding messages being transmitted through CAN, are the same. The schedule analyzer implemented in each ECU analyzes the CAN bus for a reconnaissance period of $recon$ CAN hyper-periods and sorts the instances of target message $m_v$ of each safety-critical task $T_v$ running in that ECU in descending order of corresponding ASP. Following this order, the $Obf_1$ will skip the corresponding instances of $T_v$ one by one in the ECU's schedule as long as the CLF property holds (Claim.~\ref{thm:clf}). For example, by observing the CAN schedule for $recon=2$ hyper-periods in Fig.~\ref{fig:schedAnalysis}, we can see the $4$-th instance of $m_v$ in a CAN hyper-period exhibits the highest ASP followed by $3$-rd, $2$-nd, and $1$-st instances of $m_v$. Accordingly, $Obf_1$ decides to skip the $4$-th and $3$-rd instances of $T_v$ in the subsequent schedules of the victim ECU as shown in Fig.~\ref{fig:obfschedAnalysis}. As an effect, at each hyper-period of the next reconnaissance period, the  $4$-th and $3$-rd instances of $m_v$ will not be present in the bus. \\
$\bullet$ \textbf{Obfuscation Strategy 2 ($Obf_2$)} - \emph{Skipping instances of the tasks with a priority higher than victim task}: $Obf_1$ cannot skip consecutive task instances beyond the {\em skip limit} set by the CLF property of the corresponding controller (see Sec.~\ref{subsec:cpsModel}). For example, the skip limit of the control task $T_v$ in Fig.~\ref{fig:obfschedAnalysis} is $2$ according to its CLF property. Therefore, $Obf_1$ cannot choose to skip the $2$-nd instance of $T_v$. Since these strategies are to be chosen locally in an ECU, it can only decide which among the tasks executing on the current ECU can be skipped to reduce ASP. 
If instances of $T_v$ cannot be skipped to respect the skip limit, as part of Obfuscation Strategy 2, the \hns module skips a control task that $(i)$ is of higher priority than $T_v$, $(ii)$ transmits a control message with ID less than $m_v$, and  $(iii)$ contributes to the attack window before the current instance of $m_v$. To elaborate, if $T_v^-$ is the higher priority task that executes before $T_v$ following the static schedule in the victim ECU, its corresponding message $m_v^-$ comes before $m_v$ in the CAN bus. If $m_v^-$ falls inside the attack window of $m_v$, skipping $T_v^-$ will reduce $m_v$'s attack window length. This reduces the ASP of a targeted message instance. In the best case, the ASP may even reduce to $0$. For example, notice the attack window of the $2$-nd instance of $m_v$ in the $2$-nd CAN hyper-period in Fig.~\ref{fig:schedAnalysis}. One of the higher-priority messages in this attack window corresponds to a task in the victim ECU having higher priority than $T_v$. Skipping this higher priority task in the ECU's schedule will reduce the attack window size to $1$ in the next reconnaissance period as shown in Fig.~\ref{fig:obfschedAnalysis}, and thereby, reduce the corresponding ASP.\\
$\bullet$ \textbf{Obfuscation Strategy 3 ($Obf_3$)} - \emph{Executing the victim task first among the same priority tasks}: As discussed earlier, any control task cannot be skipped beyond the skip limit set by its CLF property. For example, $Obf_2$ policy has skipped executing one instance of $T_v^-$, say $T_{v_2}^-$, in victim ECU whose corresponding message instance $m_v^-$ belongs to the attack window of $m_v$'s $2$-nd instance. Considering the skip limit for $T_v^-$ is $1$, $Obf_2$ cannot skip the particular instance of $T_v^-$, say $T_{v_1}^-$, whose corresponding message instance $m_v^-$ falls into the attack window of $m_v$'s $1$-st instance (see the $2$-nd CAN hyper-period in Fig.~\ref{fig:obfschedAnalysis}). As can be seen in Fig.~\ref{fig:obfschedAnalysis}, the two instances $T_{v_1}^-$ and $T_{v_2}^-$ of $T_v^-$ come sequentially in the ECU's schedule. Either of $T_{v_1}^-$ and $T_{v_2}^-$ can be skipped, but not both, according to $T_v^-$'s CLF property.  	
	In such a scenario, the \hns framework opts for another obfuscation strategy $Obf_3$.
	$Obf_3$ figures out the set of tasks $T_v^<$ (includes $T_v$) in the ECU i) having the same priority as $T_v$, and also, ii) belonging to the attack window of $m_v$ in the CAN bus. At certain time instances when the scheduler in ECU has to pick one among the tasks in $T_v^<$, $Obf_3$ dictates it to select the victim task $T_v$ first during the subsequent reconnaissance period.
  This ensures that the set of higher priority messages $m_{v^<}$ corresponding to the tasks in $T_v^<$ will come after $m_v$ in the next reconnaissance period.
  If the cardinality of the set $T_v^<$ (and $m_v^<$) is $\norm{T_v^<}$ ($\norm{m_v^<}$), the application of obfuscation strategy 3 will reduce the attack window of $m_v$ by $\norm{m_v^<}$, and thereby, reduce the corresponding ASP. In Fig.~\ref{fig:obfschedAnalysis}, consider that the yellow and green patterned and the grey tasks have the same priority in the victim ECU. However, only the message corresponding to the green patterned task belongs to the attack window of the $1$-st instance of $m_v$ (in the $2$-nd CAN hyper-period). We can see that executing $T_v$ before the green patterned task in the ECU's schedule reduces the attack window of $m_v$'s $1$-st instance from $2$ to $1$. \\
\textbf{\aawsos - Defends as well as Detects SBA attempts}: 
Analyzing the CAN traffic during the current reconnaissance period, the attacker finds out the target instances of the victim message with higher ASP i.e. non-zero attack window length. The obfuscation strategies discussed above ensure the reduction of ASP of each such probable target message instance by $hiding$ their timing information in the next reconnaissance period. The \aawsos strategies either skip or relocate the target message instances. An attacker may attempt to launch the SBA in the next reconnaissance period by injecting a message with the same ID as the victim message during the attack window. The \aawsos strategies invalidate the attacker's calculation of the precise arrival time of the target message instances, and thereby, quash the SBA attempt. Consequently, \aawsos strategies help in $seeking$ the attack attempt. For example, analyzing the current reconnaissance period of $recon$=2 CAN hyper-periods in Fig.~\ref{fig:schedAnalysis}, the attacker may target the $4$-th instance of $m_v$ to launch SBA. It fabricates $atk$ message with the same ID as $m_v$ and injects it into the CAN bus during the attack window of $m_v$'s $4$-th instance in the next reconnaissance period. However, since $Obf_1$ has skipped the corresponding execution of $T_v$ in the victim ECU, ideally the $4$-th instance of $m_v$ should not be present in the CAN bus. Observing a message with the same ID as $m_v$ in the CAN bus during the same time slot, the \hns module will perceive it as coming from a compromised ECU and raise an alarm to indicate a bus-off attempt.   

\subsection{Attack Success Probability under \hns} 
\label{subsec:asp}
To analyze how effectively the proposed \aawsos strategy defends against SBAs, we evaluate the attack success probability (ASP) on a CAN hyper-period with and without this strategy. As explained in Sec.~\ref{subsec: schedAnalysis}, launching an SBA requires a detailed timing analysis of CAN traffic for a certain number of hyper-periods which we call \emph{reconnaissance period}. Let us consider that the reconnaissance period for an attacker is $recon$ number of CAN hyper-periods (i.e. $\{H_1, H_2\cdots H_{recon}\}$).
Based on the CAN transmission schedule, we divide each CAN hyper-period into multiple slots. Consider there are $n_i$ transmissions with ID $i$ for a task $T_i$ where $i\in IDset$. Here, $IDset$ denotes the set of message IDs corresponding to control tasks transmitted over every CAN hyper-period of length $H$.
This means the message with ID $i$ uses $n_i$ number of slots for transmissions in $H$ duration. If the $j$-th slot in $H$ is used to transmit the message with ID $i$, then the length of the $j$-th slot equals the time required to transmit the message ID $i$. The total number of slots for transmission of all message IDs in a single CAN hyper-period is $J = \sum\limits_{\forall i\in IDset}n_i$. Now, to quantify an SBA effort, we utilize this concept of slots. Let us consider $T_v$ as the victim task that transmits the message $m_v$ with ID $v$. As explained in Sec.~\ref{subsec: schedAnalysis}, there are two desired events that the attacker observes during the reconnaissance period for launching a successful SBA. \emph{First}, the attacker needs to check whether the victim message $m_v$ arrives at $j$-th slot of every hyper-period during reconnaissance (i.e. during $\{H_1, H_2,\cdots, H_{recon}\}$). \emph{Next}, the attacker observes whether there is \emph{non-zero} number of higher priority messages before the victim message $m_v$ at every hyper-period during reconnaissance. This helps the attacker decide whether or not to launch SBA on $m_v$ at $j$-th slot.
We define a function $ct:IDset\times [1,n]\mapsto {0,1}$ as follows.$ct(v,j)= 1 \text{if $T_v$ transmits ($m_v$) at $j$-th slot}$ and $ct=
    0 \text{if $T_v$ does not transmit at $j$-th slot}$.
Let us consider that the number of higher priority messages preceding the target/victim ID $v$ at $j$-th slot is $n_{v,j}$.
To quantify the chance of launching a successful SBA at $j$-th slot, we can measure the ratio between the number of slots/transmissions with higher priority and the total number of transmissions between the current and last instance of $m_v$. If we denote the count of the total number of \textbf{\em t}ransmissions \textbf{\em b}etween the current and last \textbf{\em i}nstances of $m_v$ at $j$-th slot as $TBI_{v,j}$ 
the \emph{attack success} at $j$-th slot can be measured as, $P(AS_j)=\sum\limits_{k=1}^{recon}\frac{ct(v,kJ+j)}{recon}\cdot \frac{n_{v,kJ+j}}{TBI_{v,kJ+j}}$. This is nothing but the average of the ratios of higher priority tasks before the victim ID to $TBI$ ($\frac{n_{v,j}}{TBI_{v,j}}$) at every $j$-th slot where it appears in every hyper-period $\in \{H_1,H_2,\cdots,H_{recon}\}$. 

A successful SBA launch event on a victim ID $m_v$ occurs when SBA can be launched at any one of the $J$ transmission slots during the reconnaissance period.
Since being successful at different slots are mutually exclusive events, the overall SBA success probability for the attacker becomes
$P(AS)= P(\bigcup\limits_{\scriptscriptstyle\forall j\in [1,recon.J]}AS_j) = \sum\limits_{\forall j\in [1,recon.J]}P(AS_j)$. 
To establish the effectiveness of our obfuscation policy, we claim the following.
\begin{claim}
Application of the \emph{Attack-aware schedule obfuscation strategies} (Sec.~\ref{subsec:obfRules}) reduces the \emph{ASP} of SBA.\hfill$\Box$
\end{claim}
\textbf{Proof:}
    As discussed in Sec.~\ref{subsec:obfRules}, \hns has $three$ obfuscation strategies $\{Obf_1,Obf_2,Obf_3\}$, either of which we apply at every victim task instance in the victim ECU i.e. at every $j$-th slot of a CAN hyper-period, whenever the victim task $m_v$ appears. Let us denote $O_i$ as the event of applying obfuscation strategy $Obf_i$ where $i\in\{1,2,3\}$.
    It is intuitive that $O_i$'s are mutually exclusive. Therefore, the probability of applying any obfuscation strategy can be calculated  
$P(Obf)=P(O_1\cup O_2\cup O_3)=\sum\limits_{i\in {1,2,3}}P(O_i)$.
The ASP under \hns can
therefore be evaluated using the following conditional probability.
\begin{align*}
P(AS|Obf)&=P(\bigcup\limits_{\scriptscriptstyle\forall j\in [1,reconJ]}(AS_j|(O_1\cup O_2 \cup O_3))\\ 
&= \sum\limits_{\substack{\scriptscriptstyle\forall j\in [1,reconJ]\\\scriptscriptstyle s.t.\ ct(v,j)=1}}\frac{(P(AS_j\cap (O_1 \cup O_2\cup O_3))}{P(O_1 \cup O_2\cup O_3)}\\
&= \sum\limits_{\substack{\scriptscriptstyle\forall j\in [1,reconJ]\\\scriptscriptstyle s.t.\ ct(v,j)=1}}\frac{\sum\limits_{\forall q\in\{\scriptscriptstyle1,2,3\}}P(AS_j| O_q)P(O_q)}{\sum\limits_{\forall q\in\{\scriptscriptstyle1,2,3\}}P(O_q)}
\end{align*}
If we apply either of the obfuscation policies at every instance of the victim  task execution in the victim ECU which transmits the victim message $m_v$ in CAN, then the denominator of the above conditional is evaluated to be $1$ (i.e. $O_1$, $O_2$ and $O_3$ are exhaustive). Hence,
\begin{align*}
    P( AS| Obf)& \approx \sum\limits_{\substack{\scriptscriptstyle\forall j\in [1,reconJ]\\\scriptscriptstyle\text{s.t. }ct(v,j)=1}}\sum\limits_{\scriptscriptstyle q\in\{\scriptscriptstyle1,2,3\}}P(AS_{j}| O_q)P(O_q)\\
    &= \sum\limits_{\substack{\scriptscriptstyle\forall j\in [1,reconJ]\\\scriptscriptstyle\text{s.t. }ct(v,j)=1}}P(AS_j|Obf) \\
    \Rightarrow P(AS_j|Obf) &\approx \sum\limits_{\scriptscriptstyle q\in\{\scriptscriptstyle1,2,3\}}P(AS_{j}| O_q)P(O_q)
\end{align*}
Now, let us observe how the probability of the \emph{attack event} under the application of the obfuscation rules, changes.
    \noindent\par $\bullet$ The $Obf_1$ strategy skips the execution of the $T_v$ instance which transmits $m_v$ at $j_v$-th slot. Therefore, we can say on application of $Obf_1$, $ct(v,j_v)=0 \Rightarrow P(AS_{j_v}|O_1)=0< P(AS_{j_v})$.
    \noindent\par $\bullet$ Similarly, as per the $Obf_2$ strategy, the execution of the higher-priority task instance $T^-_{v}$ is skipped as it is executed before the current instance of $T_v$ and contributes to the attack window before $m_v$ at $j_v$-th slot. In the best case, the attack window length can become $0$. Therefore, the application of $Obf_2$ reduces the attack window length from $n_{v,j_v}$ to $n_{v,j_v}\prime$ such that $n_{v,j_v}\prime < n_{v,j_v}$. This implies $P(AS_{j_v}|O_2)< P(AS_{j_{v}})$.
    \noindent\par $\bullet$ Finally, $Obf_3$ strategy
    shifts the execution of the job-instances of $T_v$ which transmits $m_v$ at $j_v$-th slot, before all the tasks of its same priority. We do so, only if the same-priority tasks in ECU-level
    transmit higher priority messages in CAN that contribute to the attack window before $m_v$ at $j_v$-th slot. Therefore, we can say the application of $Obf_3$ reduces the number of preceding higher-priority messages. Say, the set of tasks that are of the same priority as $T_v$ and have contributed to the attack window before $m_v$ in CAN's $j_v$-th slot, is denoted as $T_v^<$. The $Obf_3$ will reduce the attack window length before $m_v$ to $n_{v,j_v}\prime = n_{v,j_v} - |T_v^<| < n_{v,j_v}$. This implies $P(AS_{j_v}|O_3)< P(AS_{j_{v}})$. 
As the \emph{attack event conditional under obfuscation policies} is less than the actual attack event probabilities without obfuscations, the ASP at any $j$-th slot is definitely less when the obfuscation strategies are applied, i.e. $P(AS_j|Obf)<P(AS_j)$. Since the overall ASP is the sum of ASPs at every $j$-th slot where the victim task arrives within the reconnaissance period, the overall ASP under our  \aawsos is also reduced. This proves the claim.
$\hfill\Box$

\noindent$\bullet$ \textbf{Comparison with \emph{typical schedule randomization} Strategies:} Notice that the attack event conditional under \hns policy is nothing but a weighted average of attack success conditionals under different obfuscation policies: 
$P(AS_j|Obf)= \sum\limits_{\scriptscriptstyle q\in\{\scriptscriptstyle1,2,3\}}P(AS_j| O_q)P(O_q)$.
As part of \hns, we apply $Obf_1$ and $Obf_2$ as our primary preferences. 
As discussed earlier, $P(AS_{j}|O_1)=0$ and $P(AS_{j}|O_2)<P(AS_{j})$. 
When we do not have the luxury to skip any more control executions in consecutive iterations to respect the \emph{skip-upper-bounds} of the closed-loops, we apply $Obf_3$ as our third choice. This means, $P(O_3)$ depends on the upper bounds of the consecutive execution skips. Note that, $P(AS_{j})$ due to the application of $Obf_3$ policy is $\frac{n_{v,j_v} - |T_v^<|}{TBI_j}$. This leads to a reduction in the ASP by $\frac{|T_v^<|}{TBI_j}$ from the situation before the schedule transformation with $Obf_3$.
\par \emph{Attack-unaware} schedule randomizations have been proposed in earlier works  in order to conceal timing side channels exposed  by processor-level task schedules~\cite{kruger2018vulnerability}. Note that w.r.t. a bus-off attacker, the most resilient schedule that may come out of such non-informed schedule randomization is basically the outcome of the application of $Obf_3$. In this case, task skipping does not happen, so $Obf_1$ and $Obf_2$ do not exist. Hence, under randomization techniques like \cite{kruger2018vulnerability}, the ASP lower bound is $\frac{n_{v,j} - |T_v^<|}{TBI_j}$.
Whereas, in the case of the \aawsos, the ASP is significantly less due to additional application of the other two 
policies (i.e. $Obf_1$ and $Obf_2$) that have the ability to reduce $P(AS_j|Obf)$ to $zero$ in suitable cases. This evidently concludes the fact that our \aawsos policy is more apt in terms of defending against SBAs compared to any other schedule randomization technique.
\begin{algorithm}[!ht]
\scriptsize
\caption{\small Control-skip-guided Attack-aware Schedule Obfuscation and Detection Algorithms }
\label{algo:hns}
\begin{algorithmic}[1]
    \Require{$H$, $recon$, $IDset$, $atkWinlist$, $TS$, $h$, $skiplist$}
    \Ensure{$ObfScheds,alarm$}
    \Function{Hide-n-Seek}{$H$, $recon$, $IDset$, $atkWinlist$, $TS$, $h$, $skiplist$}
    \For{$every\ reconnaissance\ period$}\label{alg:mainfor}
    \State $ObfScheds\gets $\textsc{Hide($H$, $recon$, $IDset$, $atkWinlist$, $TS$, $h$, $skiplist$)}\label{alg:mainhide}
    \State $alarm \gets \Call{Seek}{H, recon, ObfScheds, skiplist}$\label{alg:mainseek}
    \If{$alarm$} \emph{take safety/security measure}\label{alg:mainalarm}
    \EndIf
    \EndFor\label{alg:mainforend}
    \State \Return $ObfScheds$, $alarm$
    \EndFunction
    \Function{Hide}{$H$, $recon$, $IDset$, $atkWinlist$, $TS$, $h$, $skiplist$}\label{alg:hide}
    \State $rpt\gets \ceil{\frac{recon\times H}{h}}$\label{alg:hiderpt}
    \State $Scheds,ObfScheds,eqPrilists\gets [\ ],[\ ],[\ ]$\label{alg:hideinit}
    \While{$rpt > 0$}\label{alg:hidewhilest}
        \State $eqPrilists, Sched\gets Scheduler(TS)$\label{alg:hideedf}
        \State $Scheds\gets Scheds.append(Sched)$\label{alg:hideedfappend}
        \State $eqPrilists\gets eqPrilists.append(eqPrilist)$\label{alg:hidecontappend}
        \State $ObfScheds\gets ObfScheds.append(Sched)$\label{alg:hideobfappend}
        \State $rpt\gets rpt-1$\label{alg:hiderpt-}
    \EndWhile
    \State sort $IDset$ in descending order of \emph{attack window} length in $atkWins$\label{alg:hideidsort}
    \For{each $id \in IDset$}\label{alg:hideforid}
        \State $atkWins\gets atkWinlist[id]$\label{alg:hideisvid}
        \State $ideqPrilist\gets eqPrilists[id]$\label{alg:hidecontid}
        \State sort $atkWins$ in descending order of slot vulnerability value\label{alg:hidesortsv}
        \State $ECUslots\gets slotMapper(atkWins.slots)$\label{alg:hideecuslot}
        \For{each $si \in ECUslots$}\label{alg:hideforslot}
            \State $ifObf_1\gets \Call{checkSkipLim}{id,si,ObfScheds,skiplist}$\label{alg:hidecheckskip}
            \If{$Obf_1$} $ObfScheds[si]\gets 0$\label{alg:hideskip1}
            \Else
            \State $idPreds, sPreds \gets \Call{getHpPreds}{ideqPrilist[id]}$\label{alg:hidepredslot}
            \State sort $idPreds, sPreds$ in descending order of start time\label{alg:hidepredslotsort}
            \State $idPred,sPred\gets idPreds[0], sPreds[idPreds[0]]$\label{alg:hidepredinit}
            \For{each $idPred\in idPreds$}\label{alg:hideallpreds}
                \State $ifObf_2\gets$ \textsc{checkSkipLim($idPred$,$ObfScheds$,$skiplist$)}\label{alg:hidecheckskippred}
                \State $sPred\gets sPreds[idPred]$\label{alg:hidecheckstorepredslot}
                \If{$ifObf_2$} $ObfScheds[sPred]\gets 0$\label{alg:hideskip2}
                \State $break$\label{alg:hidefoundskippredbreak}
                \EndIf
            \EndFor
            \If{ $\neg(ifObf_1 \wedge ifObf_2)$}\label{alg:hideifnot12}       \Comment{$Obf_3$}
            \State $ObfScheds \gets \Call{shiftVic}{ObfScheds,si,sPreds}$\label{alg:hideskip3}
            \EndIf
            \EndIf
        \EndFor
    \EndFor
    \State \Return $ObfScheds$\label{alg:hideobfret}
    \EndFunction
    \Function{Seek}{$H,recon,ObfScheds,skiplist$}\label{alg:seek}
    \State $si,alarm\gets 0,0$\label{alg:seekinit}
        \For{$si < recon.J$}\label{alg:seekfor}
        \State $ECUSlot \gets slotMapper(si)$ \label{alg:slotMapperInSeek}
        \If{$ObfScheds[si]\neq ECUSlot$}\label{alg:seekif}
        \State $alarm\gets 1$\label{alg:seekalarm}
        \EndIf
        \EndFor
        \Return $alarm$\label{alg:seekret}
    \EndFunction
\end{algorithmic}
\end{algorithm}

\subsection{Algorithmic Framework for \emph{\hns} }
\label{subsec:hnsalgo}
We propose an algorithmic framework of the proposed \hns strategy that should be locally implemented on every ECU connected to the CAN bus. We assume the execution of the task set $TS$ follows a static schedule in an ECU. As a prerequisite to running the algorithm, we consider that a thorough analysis of CAN traffic for $recon$ number of CAN hyper-periods is done following the state-of-the-art~\cite{hounsinou2021vulnerability} methodologies. Following is the information available from this analysis:
$(i)$ the length of CAN hyper-period $H$ and the minimum number of CAN hyper-periods required for reconnaissance $recon$; $(ii)$ the set of control message IDs transmitted via the CAN bus $IDset$; $(iii)$ The ID-wise list of attack windows (containing the consecutive higher priority transmission slots) appearing in front of any instance of a message ID generated from the current ECU ($atkWinlist$). With these as inputs, in Algo.~\ref{algo:hns}, we make two procedure calls $(i)$ $\Call{Hide()}{}$ that decide the obfuscation policies for each of the safety-critical control task instances running in the current ECU (line~\ref{alg:mainhide}); $(ii)$ $\Call{Seek()}{}$ that seeks for the presence of a bus-off attacker in CAN bus (line~\ref{alg:mainseek}).

\par In each call for every reconnaissance period, the $\Call{Hide()}{}$ procedure is run for $rpt$ number of ECU hyper-periods (lines~\ref{alg:hide}-\ref{alg:hideobfret}). Here $rpt$ is calculated in order to cover the whole reconnaissance period i.e. $recon\times H$ (see line~\ref{alg:hiderpt}). Since the task set $TS$ is static, by observing the task schedule in the current hyper-periods, we can derive the task execution schedule for the coming $rpt$ number of ECU hyper-periods. We store this extended schedule in $Scheds$ (line~\ref{alg:hideedfappend}). While deriving this schedule, we also find out the list of equal priority tasks available with every control task instance at the same time in the ECU processor.
This is stored in $eqPrilists$ as shown in line~\ref{alg:hidecontappend}. An array $ObfScheds$ is created to store the obfuscated schedules for the next reconnaissance period and initialized with the derived extended schedule (line~\ref{alg:hideobfappend}). Next, we sort the $IDset$ in descending order of the attack window lengths of each ID as stored in $atkWinlist$ (see line~\ref{alg:hideidsort}). This enables us to decide the obfuscation policies for all the control message IDs, starting from the most vulnerable ID in the $IDset$ (in line~\ref{alg:hideforid}-\ref{alg:hideobfret}). Note that, ASP increases with attack window length (Sec.~\ref{subsec: schedAnalysis}).
\par We collect the lists of attack windows and the list of equal priority task information in $atkWins$ and $ideqPrilist$ respectively for each victim ID in $IDset$ (see lines~\ref{alg:hideisvid},~\ref{alg:hidecontid}). In line~\ref{alg:hidesortsv}, the transmission instances of the current ID, stored in $atkWins$, are sorted in descending order of the attack window length. Thereafter, we find the ECU task execution slots corresponding to these sorted CAN transmission instances which are then stored in $Ecuslots$ maintaining the order (refer to line~\ref{alg:hideecuslot}). A subroutine $\Call{slotMapper}{}$ is used here that observes the relative order of the message transmissions through the CAN bus and finds out the corresponding task execution timeslot in the ECU. Having these $ECUslots$ in the descending order of attack window length, we then decide which policy to apply on which one in lines~\ref{alg:hideforslot}-\ref{alg:hideskip3}.
\par As discussed in Sec.~\ref{subsec:obfRules}, we primarily decide to apply $Obf_1$ at the most attackable task instance. For this, we must check whether skipping at the current task execution slot does not violate the mandated skip limit for the control task. We check this using a subroutine $\Call{checkSkipLim()}{}$ that takes the task ID, the skip limit of the task ID from $skiplist$ (\emph{skip limits} of every control task running in this ECU is stored in $skiplist$), and the obfuscated schedule as inputs and returns a boolean decision to denote whether a skip at the current time slot is permitted for the input task ID. For a certain ECU timeslot of the current task, we store the result in a boolean variable $ifObf_1$ in line~\ref{alg:hidecheckskip}. When $ifObf_1$ is $true$, we skip the task execution in the current timeslot and store a $0$ to denote an idle period at the current timeslot of $ObfScheds$ (line~\ref{alg:hideskip1}). If skipping at the current task instance is not permissible then we go for $Obf_2$. The $ideqPrilist$ contains the list of tasks that are of equal priority at the currently considered execution slot of a task as per the ECU schedule. A subroutine $\Call{getHpPreds()}{}$ is designed that takes this list as input and returns a list of tasks (along with their corresponding execution slots) that contributes to the attack window of the message corresponding to the currently considered victim task instance. These IDs and their execution slots are stored in $idPreds$ and $sPreds$ for further analysis (refer to line~\ref{alg:hidepredslot}). These lists are sorted in descending order of the start time of the execution slots (in line~\ref{alg:hidepredslotsort}). This helps us start the analysis for the application of $Obf_2$ with the task that transmits a higher priority message at the nearest slot before the current instance of the victim message. Starting from this, for every task in the $idPreds$ set, we check whether the skip limit of that task supports an execution skip at the current time slot $sPred$ (see line~\ref{alg:hideallpreds}). This decision is updated in the decision variable $Obf_2$ in line~\ref{alg:hidecheckskippred}. On finding such a task, we stop checking further in $idPreds$ and skip the execution of the chosen task at the $sPred$ slot by storing a $0$ in the $sPred$ slot of the obfuscated schedule array $ObfScheds$ (see line~\ref{alg:hideskip2}). When both obfuscation policies $Obf_1$ and $Obf_2$ are not applicable, we apply $Obf_3$ on the current task schedule (see line~\ref{alg:hideifnot12}). The subroutine $\Call{shiftVic}{}$ is designed to update the schedule with a valid schedule following certain scheduling algorithms (eg. EDF, RM) such that the attack window size is reduced. We apply $Obf_3$ and update the obfuscated schedule array $ObfSched$ (see line~\ref{alg:hideskip3}). After finalizing the strategies for each of the instances of each ID, we finally return the  obfuscated schedule as output in line~\ref{alg:hideobfret}.
\par The $\Call{Seek()}{}$ procedure
is designed to raise an alarm in the presence of an attacker in the CAN bus (lines~\ref{alg:seek}-\ref{alg:seekret}). We run this method for all ECU task execution slots spanning through the reconnaissance window (see line~\ref{alg:seekfor}).  As shown in line~\ref{alg:slotMapperInSeek}, we first check whether a control task is transmitting in the current slot. This finding is then verified against the obfuscated schedule of its corresponding task execution slot (line~\ref{alg:seekif}). If an instance of a control message ID appears in the bus while the decided obfuscation strategy for this message instance, mandates skipping of the corresponding task instance, then we set and return the $alarm$ to notify a schedule-based attack attempt (lines~\ref{alg:seekalarm}). 
\par The two procedures $\Call{Hide()}{}$ and $\Call{Seek()}{}$ are deployed in each ECU and run for every reconnaissance period (lines~\ref{alg:mainfor}-\ref{alg:mainforend}). The $\Call{Hide()}{}$ procedure uses the data analyzed from the CAN traffic in the current reconnaissance period and decides the obfuscation strategies for the ECU schedules of the next reconnaissance period. It returns the obfuscated schedule $ObfScheds$ (line~\ref{alg:mainhide}) following which the tasks are scheduled in ECU during the next reconnaissance period. 
On finding a message corresponding to a skipped task execution, the $\Call{Seek()}{}$ procedure  raises an alarm notifying a bus-off attack attempt (line~\ref{alg:mainseek}). The system is designed to take certain safety and security measures in such a situation in order to refrain from any adverse effects of the attack (line~\ref{alg:mainalarm}). This cyclically goes on for all the reconnaissance periods. 
%

\begin{figure*}[!ht]
    \begin{subfigure}[b]{0.32\linewidth}
        \centering
        \includegraphics[width=\textwidth]{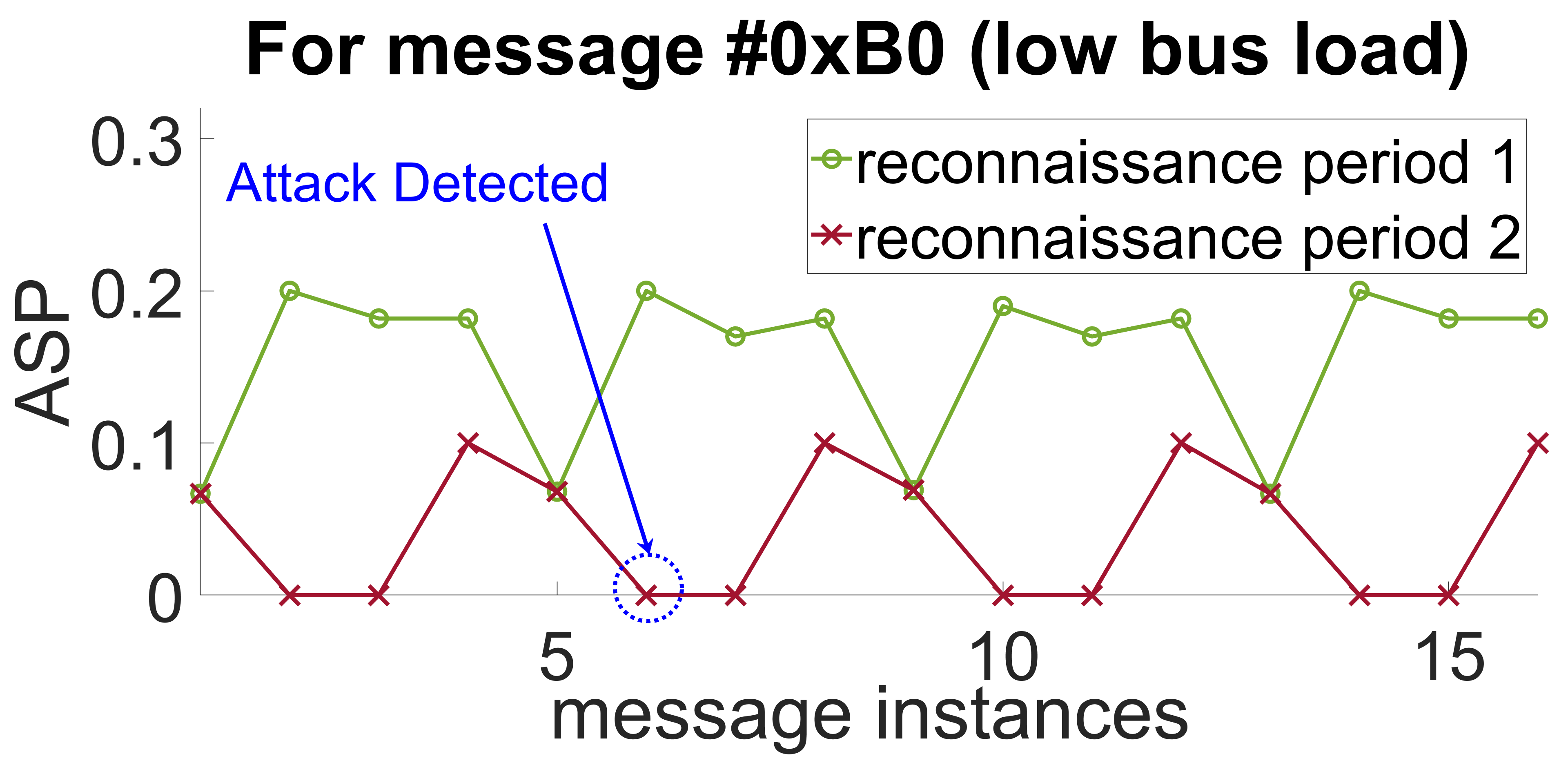}
        \caption{}
        \label{fig:LowB0}
    \end{subfigure}
    \hspace{1 mm}
    \begin{subfigure}[b]{0.32\linewidth}
        \centering
        \includegraphics[width=\textwidth]{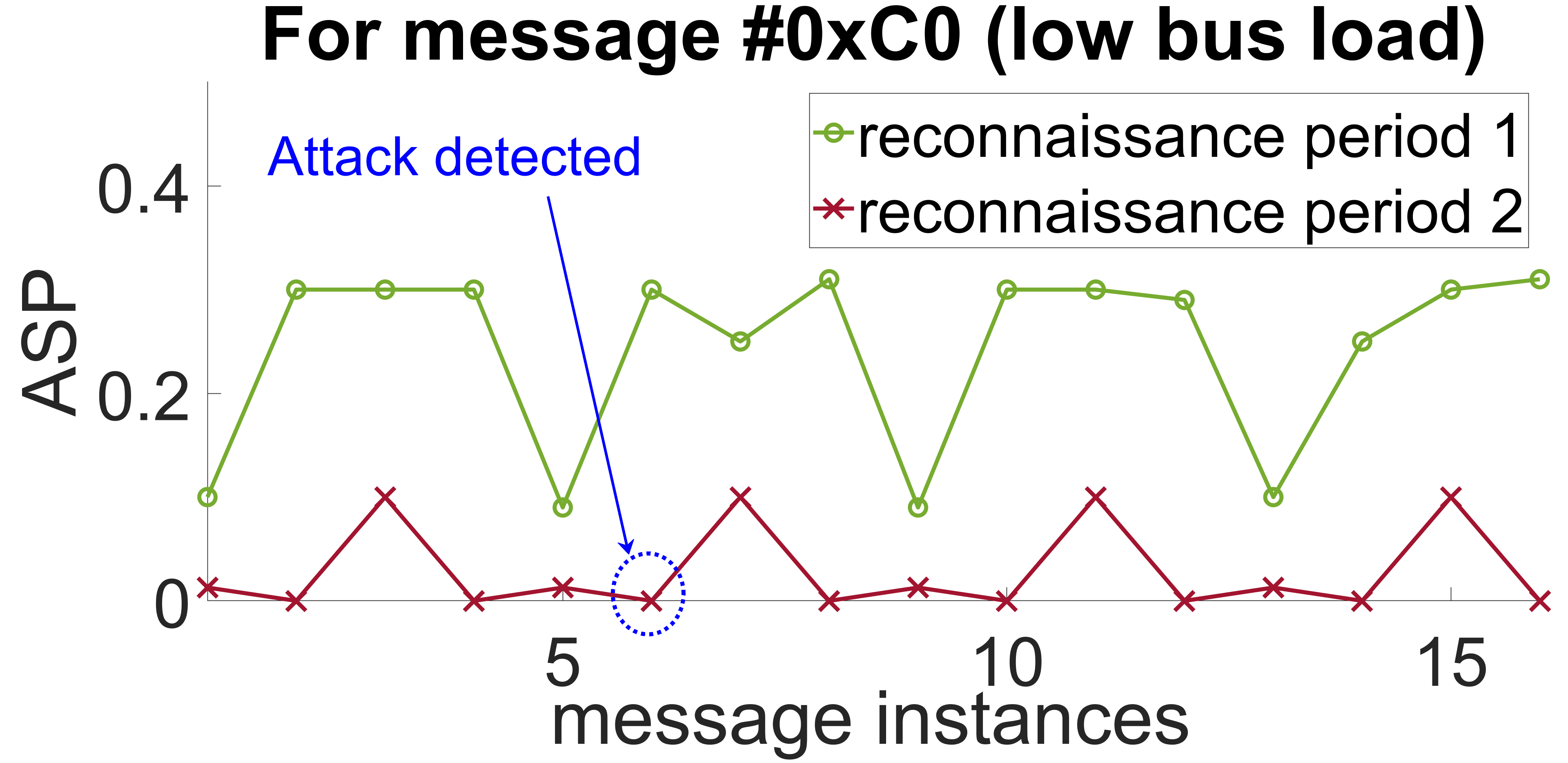}
        \caption{}
        \label{fig:LowC0}
    \end{subfigure}
    \hspace{1 mm}
    \begin{subfigure}[b]{0.32\linewidth}
        \centering
        \includegraphics[width=\textwidth]{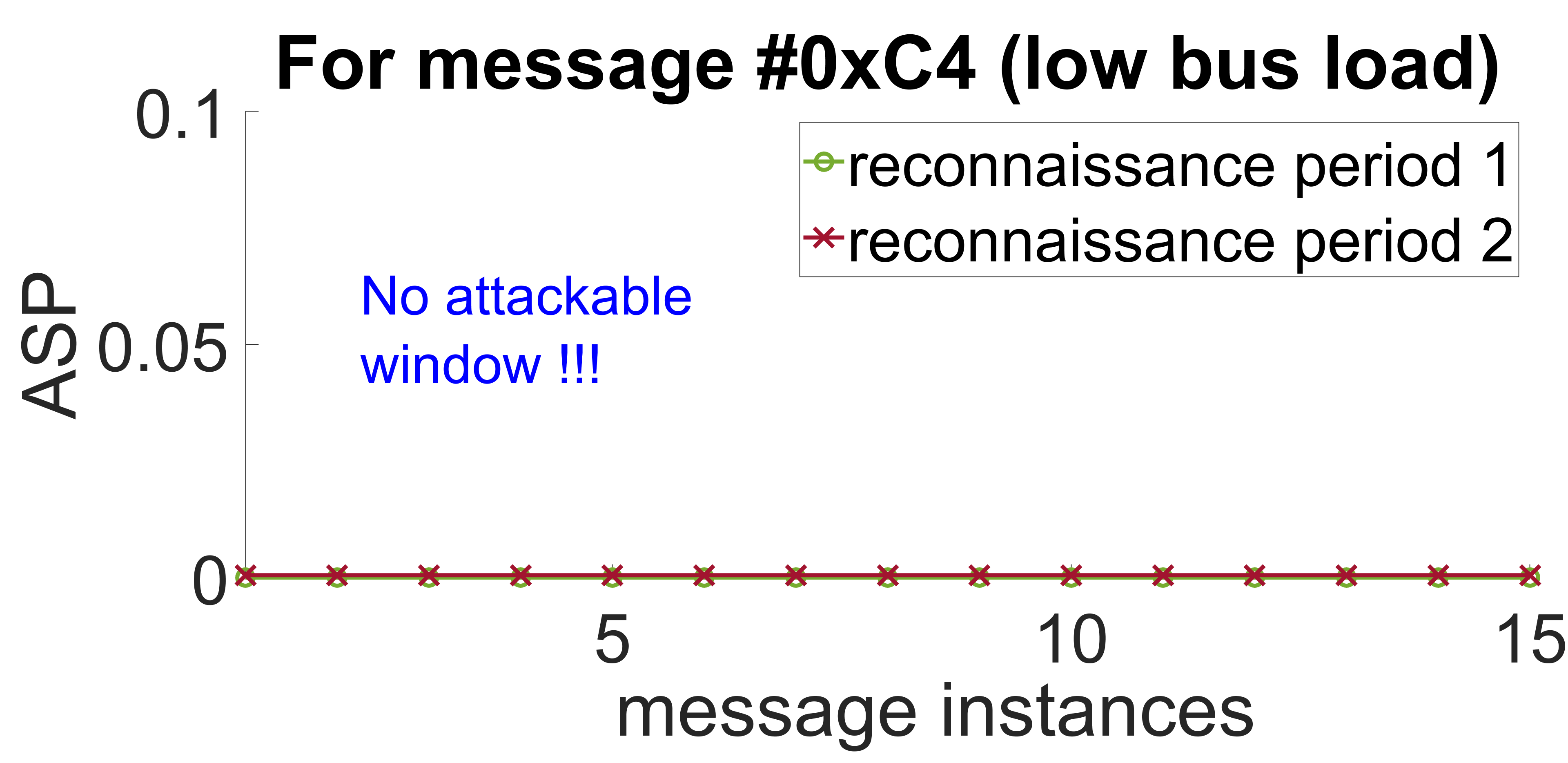}
        \caption{}
        \label{fig:LowC4}
    \end{subfigure}
    \begin{subfigure}[b]{0.32\linewidth}
        \centering
        \includegraphics[width=\textwidth]{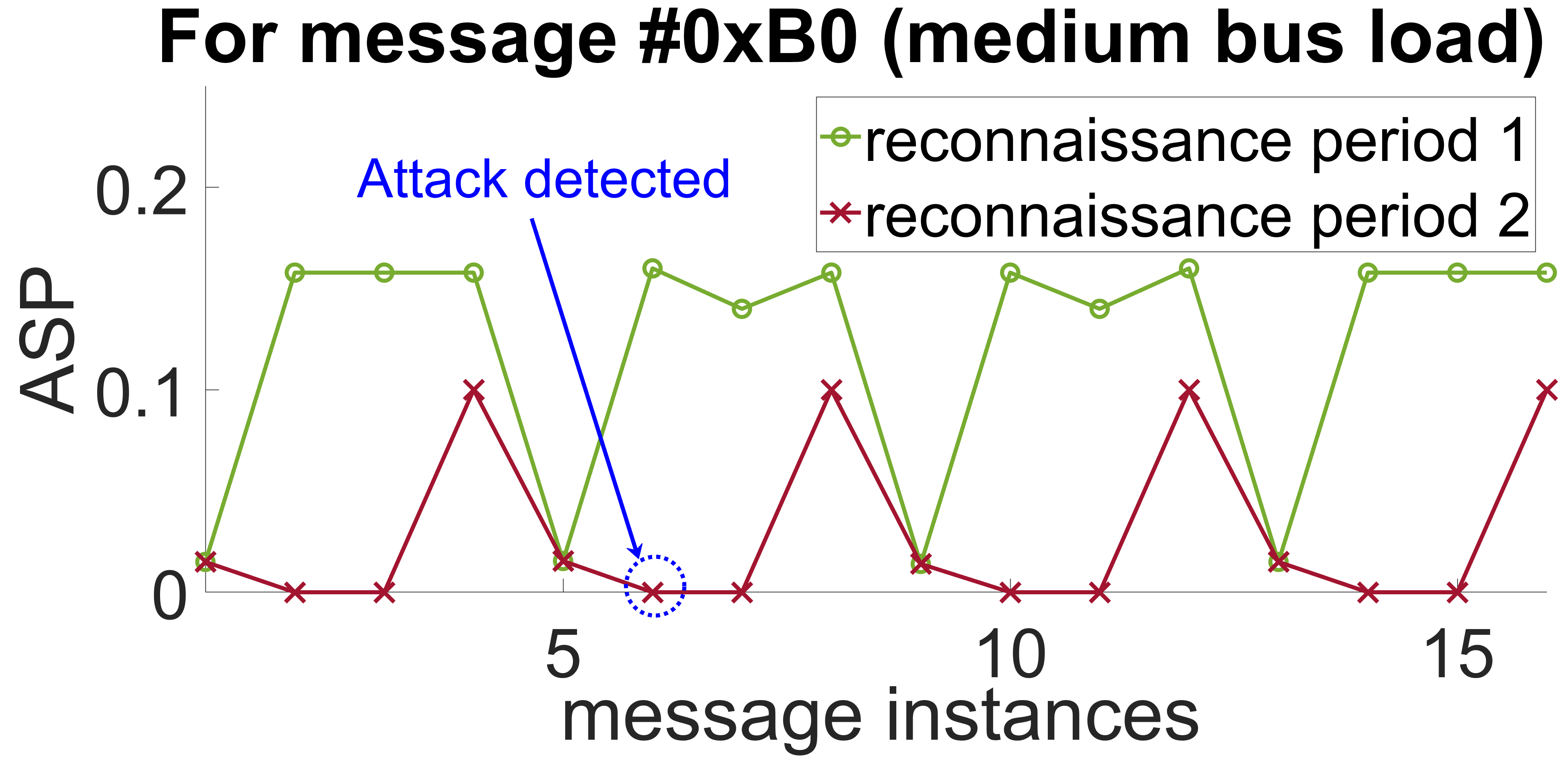}
        \caption{}
        \label{fig:mediumB0}
    \end{subfigure}
    \hspace{1 mm}
    \begin{subfigure}[b]{0.32\linewidth}
        \centering
        \includegraphics[width=\textwidth]{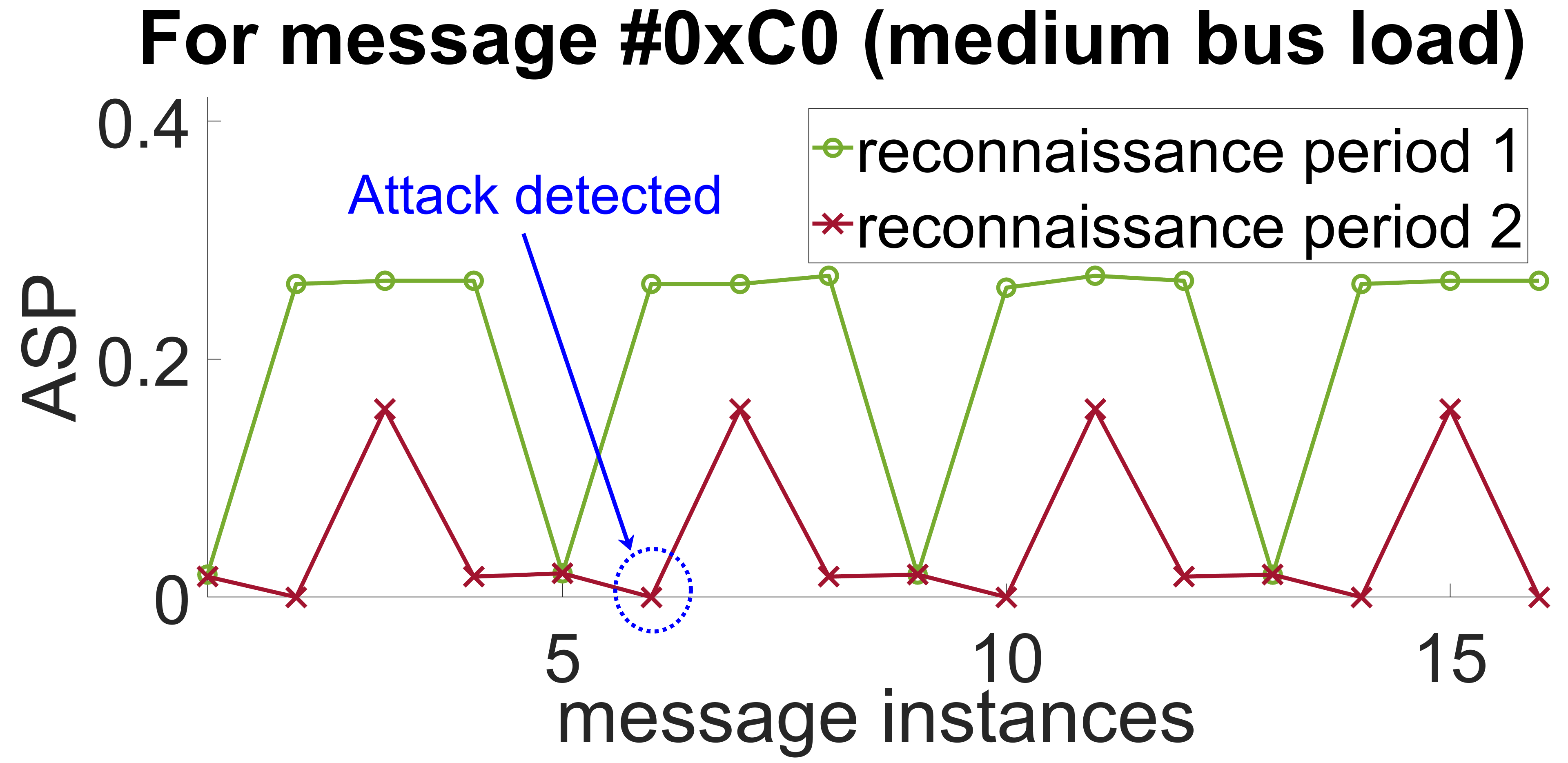}
        \caption{}
        \label{fig:mediumC0}
    \end{subfigure}
    \hspace{1 mm}
    \begin{subfigure}[b]{0.32\linewidth}
        \centering
        \includegraphics[width=\textwidth]{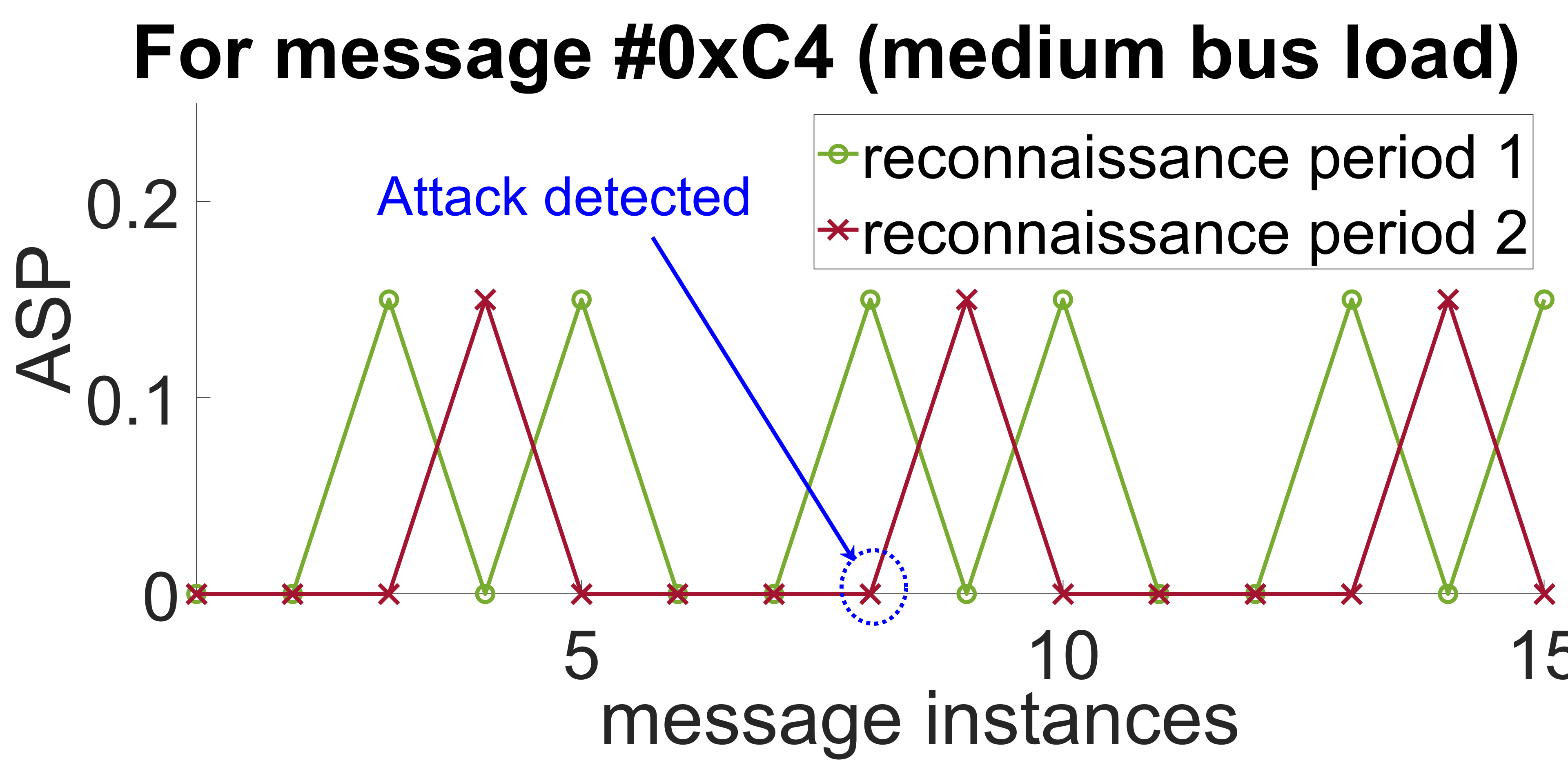}
        \caption{}
        \label{fig:mediumC4 }
    \end{subfigure}
    \begin{subfigure}[b]{0.32\linewidth}
        \centering
        \includegraphics[width=\textwidth]{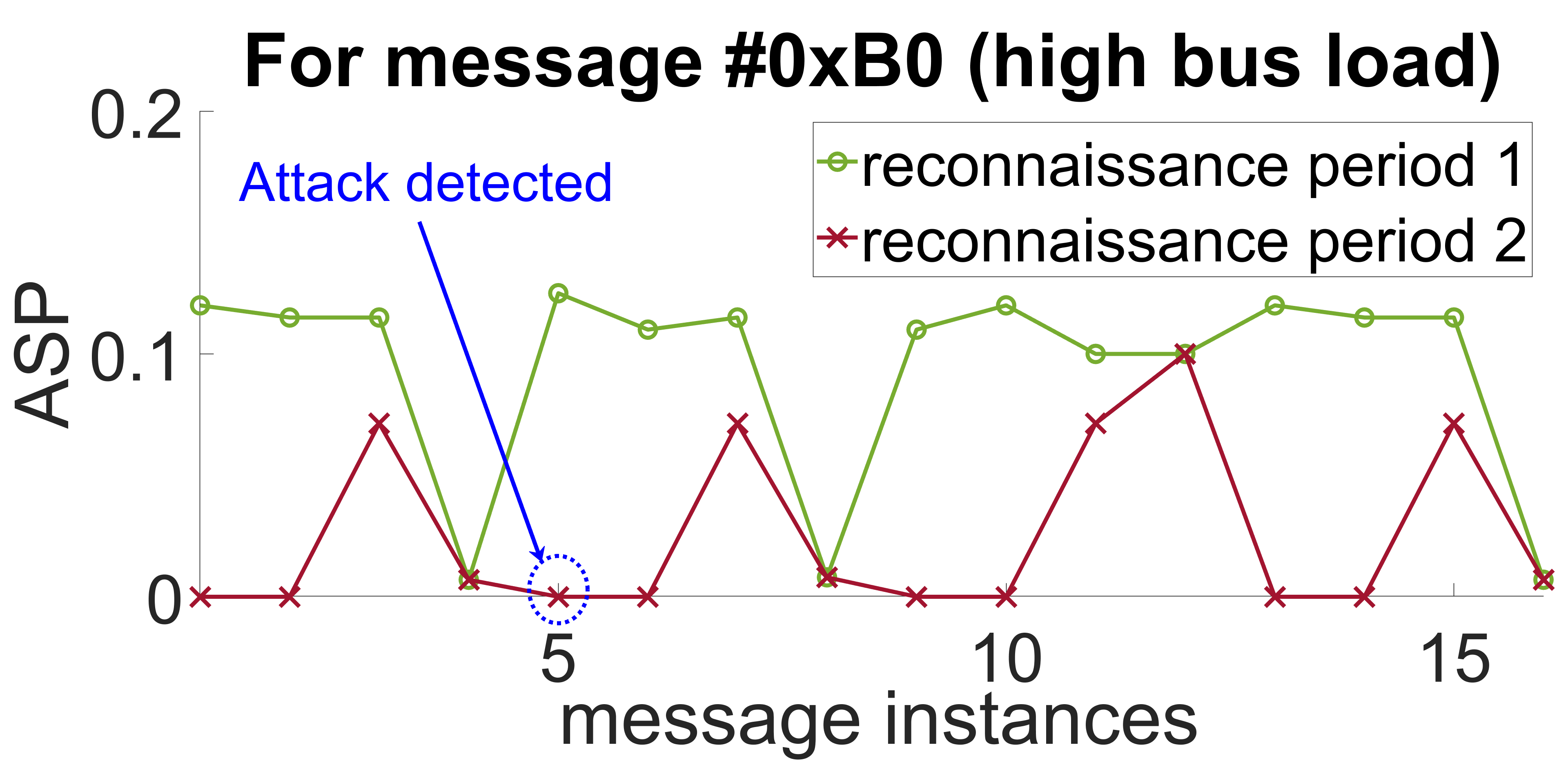}
        \caption{}
        \label{fig:HighB0}
    \end{subfigure}
    \hspace{1 mm}
    \begin{subfigure}[b]{0.32\linewidth}
        \centering
        \includegraphics[width=\textwidth]{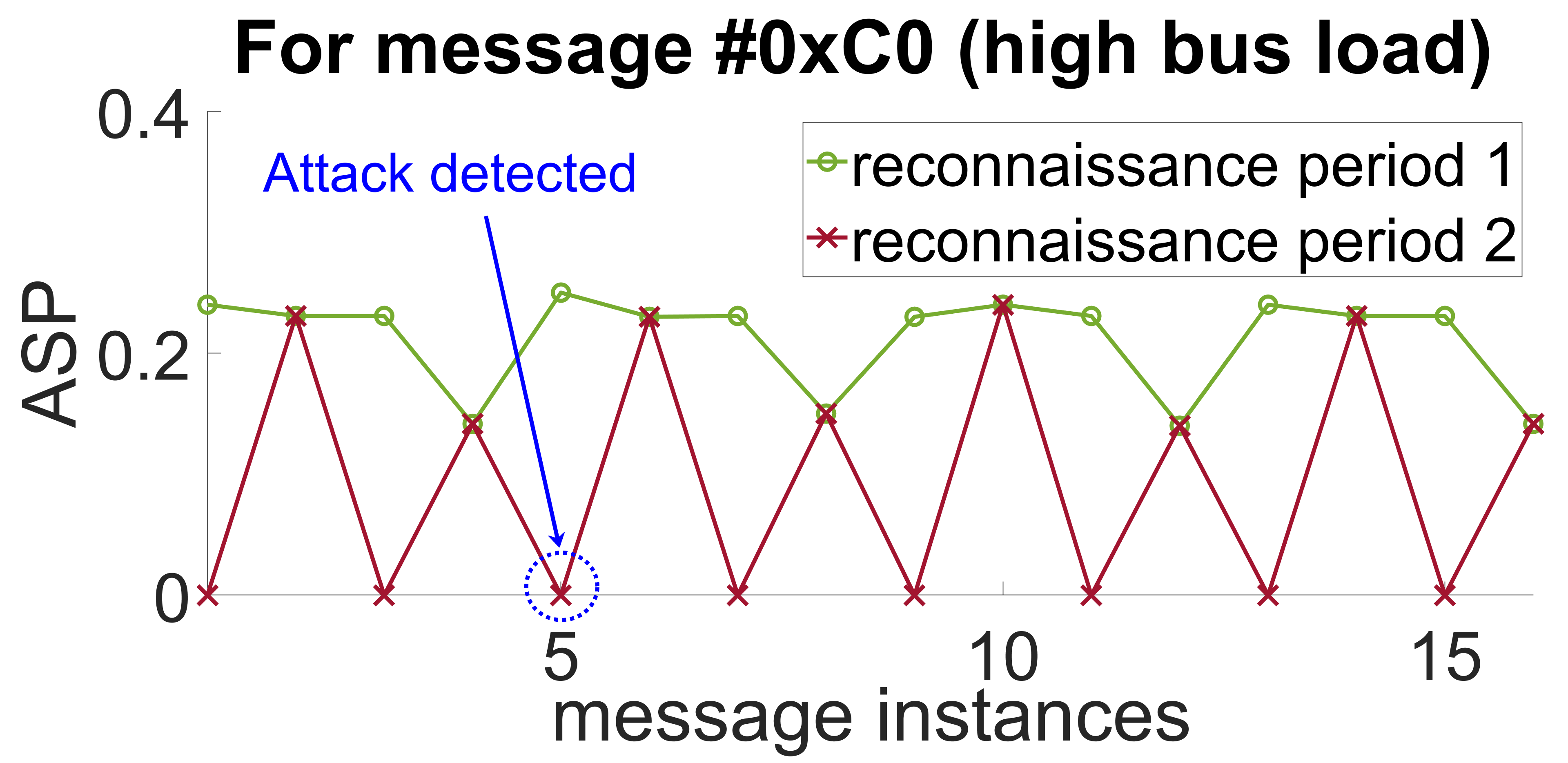}
        \caption{}
        \label{fig:HighC0}
    \end{subfigure}
    \hspace{1 mm}
    \begin{subfigure}[b]{0.32\linewidth}
        \centering
        \includegraphics[width=\textwidth]{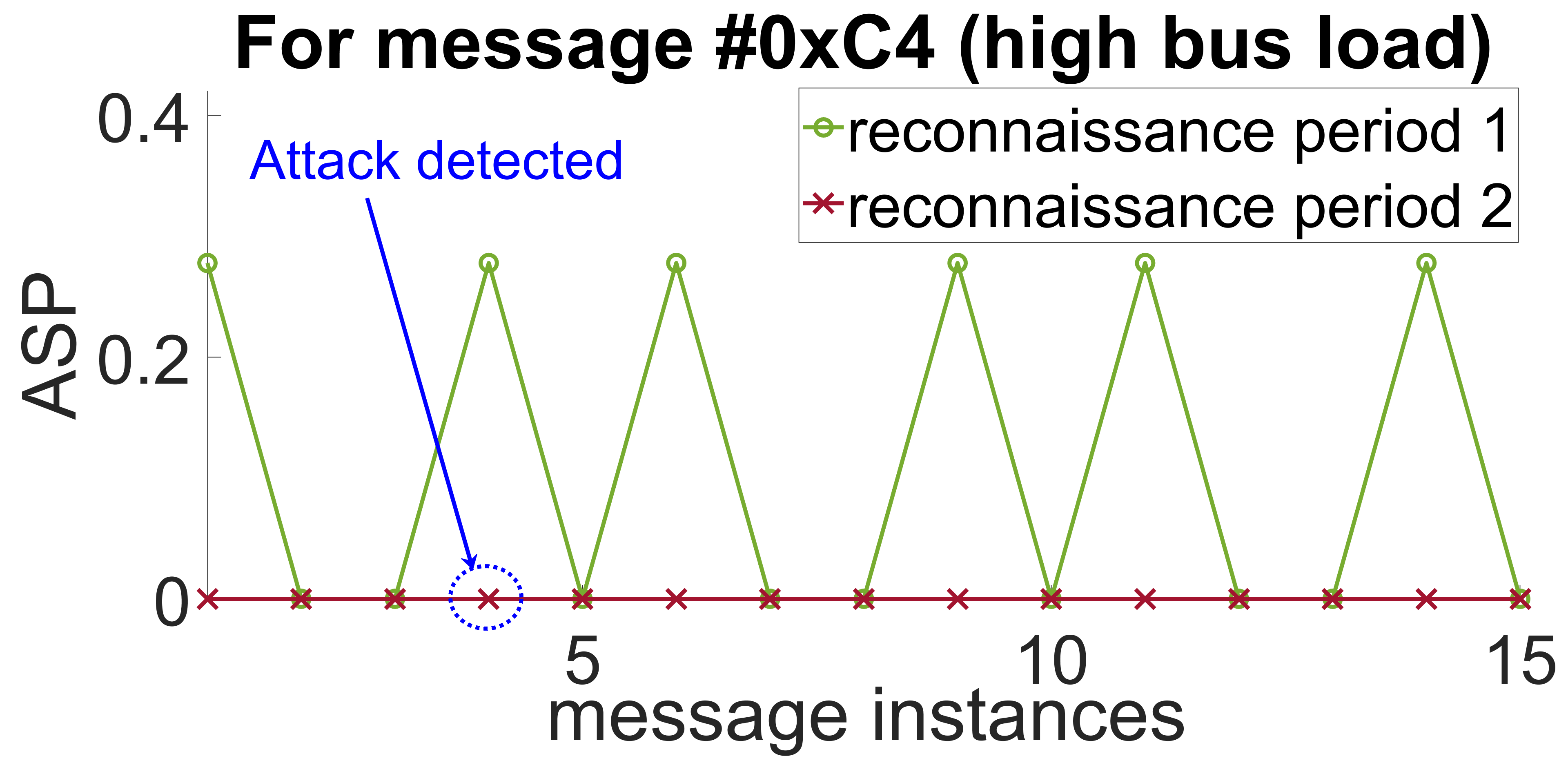}
        \caption{}
        \label{fig:HighC4}
    \end{subfigure}
    \caption{ASP analysis under schedule obfuscation.}
    \vspace{-4mm}
    \label{fig:ASP Analysis}
\end{figure*}
\section{Experimental Results} 
\label{sec:results}
We demonstrate the efficacy of the proposed \hns algorithm on practical CAN traffic under different busloads. For this, we consider a set of control tasks that are implemented across two ECUs. We use Infineon Tricore-397 and Tricore-234 ECUs for this purpose. An automotive plant is emulated in a HIL setup which is in a closed loop with these control tasks. ETAS Labcar real-time simulator is used for this purpose. The plant and the controllers communicate via the CAN bus operating with 250kbps bandwidth. The control tasks and the \hns\ algorithm are implemented and run in different cores of ECUs. A CAN schedule analysis algorithm is run in the same core as the \hns\ algorithm in order to observe all transmissions through the CAN bus. 
We enlist the detailed specifications of the tasks running in each of the ECUs and their corresponding messages in Tab.~\ref{tab:taskSet}. 

\begin{wraptable}[9]{l}{0.54\columnwidth}
\centering
\vspace{-3mm}
\caption{\footnotesize Task Setup}
\label{tab:taskSet}
\scriptsize{
\scalebox{0.85}{
\begin{tabular}{|l|l|l|}
\hline
\multicolumn{1}{|c|}{ECU} &
  \multicolumn{1}{c|}{Task Type} &
  \multicolumn{1}{c|}{Message ID} \\ \hline
\multirow{2}{*}{\begin{tabular}[c]{@{}l@{}}ECU 1\\ (TC397)\end{tabular}} &
  Control Tasks &
  \begin{tabular}[c]{@{}l@{}}0xB0, 0xC0\\ 0xC4\end{tabular} \\ \cline{2-3} 
 &
  \begin{tabular}[c]{@{}l@{}}Other Tasks\\ (* aperiodic tasks)\end{tabular} &
  \begin{tabular}[c]{@{}l@{}}0xA0, 0xB1,\\ 0xA3, 0xB3,\\ 0x$163^*$\end{tabular} \\ \hline
\multirow{2}{*}{\begin{tabular}[c]{@{}l@{}}ECU 2\\ (TC234)\end{tabular}} &
  Control Tasks &
  \begin{tabular}[c]{@{}l@{}}0x230, 0x323\\ 0x90\end{tabular} \\ \cline{2-3} 
 &
  \begin{tabular}[c]{@{}l@{}}Other Tasks\\ (* aperiodic tasks)\end{tabular} &
  \begin{tabular}[c]{@{}l@{}}0x98, 0x500, \\ 0xB2, 0xC2,\\ 0xB5, 0xC5,\\ 0x$565^*$, 0x$783^*$\end{tabular} \\ \hline
\end{tabular}
}}
\end{wraptable}
We schedule the periodic control tasks in their corresponding ECUs following an EDF schedule with $60$ ms and $30$ ms ECU hyper-periods. Whereas the CAN hyper-period is observed to be $\approx 200$ms. We consider a 3 CAN hyper period long reconnaissance period of length $\approx 600$ms. The proposed \hns\ algorithm runs for all control tasks in $ECU1$. As part of the schedule analysis, a CAN data logger is implemented to observe all messages as they appear in the CAN bus to estimate the possible attack windows for the control tasks running in this ECU during the reconnaissance period. Based on these estimations, \hns\ decides which obfuscation rules to apply on which instances of the ECU task.
\par  \noindent $\bullet$ \textit{ Evaluation of the Proposed Obfuscation Policies: } Fig.~\ref{fig:ASP Analysis} demonstrates such observed ASPs at every task instance when the tasks are scheduled with a static EDF schedule (the $circled$ $green$ plots) and the resulting ASPs on the deployment of suitable policies at every task instance in the subsequent reconnaissance period (red plots with $cross$ markers). The x-axis in Fig.~\ref{fig:ASP Analysis} denotes the relative slot index in every reconnaissance period and their instance-wise ASPs at the y-axis and their corresponding relative slot index in the x-axis.
We plot the ASP of the current reconnaissance period and the next reconnaissance period in the same plot for comparative visualization. The reductions in ASPs caused by \aawsos\ application can be easily gauged from the plots. The success of a bus-off attack highly relies on the traffic load in the bus, because with more traffic there can be longer attack windows and a better chance of going to bus-off from error passive mode due to passive error regeneration~\cite{serag2021exposing}. Therefore we report our results for 3 possible busload conditions i.e. under high ($75$\%), medium ($55$\%), and low ($25$\%) busload. As can be seen in Fig.~\ref{fig:HighB0}, $Obf_1$ is applied at those execution instances of task $T_{B0}$ (the task responsible for the transmission of message ID 0xB0) in the 2nd reconnaissance period where maximum ASPs ($\approx 0.12$) are observed under a static task schedule during the 1st reconnaissance period (i.e. at the execution slot number $2,5,6,9,10,13$ and $14$ for $T_{B0}$). This results in zero ASPs at their corresponding CAN transmission slots and leads to detection of any bus-off attack attempted during those periods (Notice in Fig.~\ref{fig:HighB0} alarm flag is raised at $5$-th slot). At $11$-th instance of $T_{B0}$ as $Obf_1$ cannot be applied in order to respect the skip limit, $Obf_2$ is applied by skipping execution of another control task $T_{C0}$ (see at $11$-th task instance in Fig.~\ref{fig:HighC0}) that contributes to the attack window of $T_{B0}$. 

At certain execution instances, \hns\ applies $Obf_3$ and reduces the ASPs where the $Obf_1$ policy can not be applied to avoid violating the skip limit of $T_{B0}$ and $Obf_2$ policy cannot be applied as none of the preceding control tasks executes and transmits messages during the attack window of $T_{B0}$ (see the 3rd, 7th and 15th instances). As can be seen in Fig.~\ref{fig:HighC4}, in case of $T_{C4}$, \hns\ applies execution skips (i.e. $Obf_1$) almost at alternate instances and is able to reduce the ASP to zero at every transmission instance of ID 0xC4. In the presence of a medium busload, the reduction in ASP is not this significant but the possibility of detection persists (see~\ref{fig:mediumC4 }). It is observable from the plots that the ASPs reduce as the busload decreases. However, it is also evident, that our obfuscation strategy reduces the ASPs under any busload. 
\par $\bullet$ {\em 
 Detection using \hns}: To demonstrate the fact that our \aawsos applies the obfuscation rules without hampering the performances of any control loops, we demonstrate the performance of the TTC closed loop under the obfuscations applied on the control task $T_{C4}$  (which transmits with ID 0xC4) during high busload situation (obfuscation decisions are shown in Fig.~\ref{fig:HighC4}). 

\begin{wrapfigure}[23]{l}{0.5\columnwidth}
    \centering
    \vspace{-4mm}
    \begin{subfigure}[b]{0.5\columnwidth}
        \includegraphics[width=\textwidth]{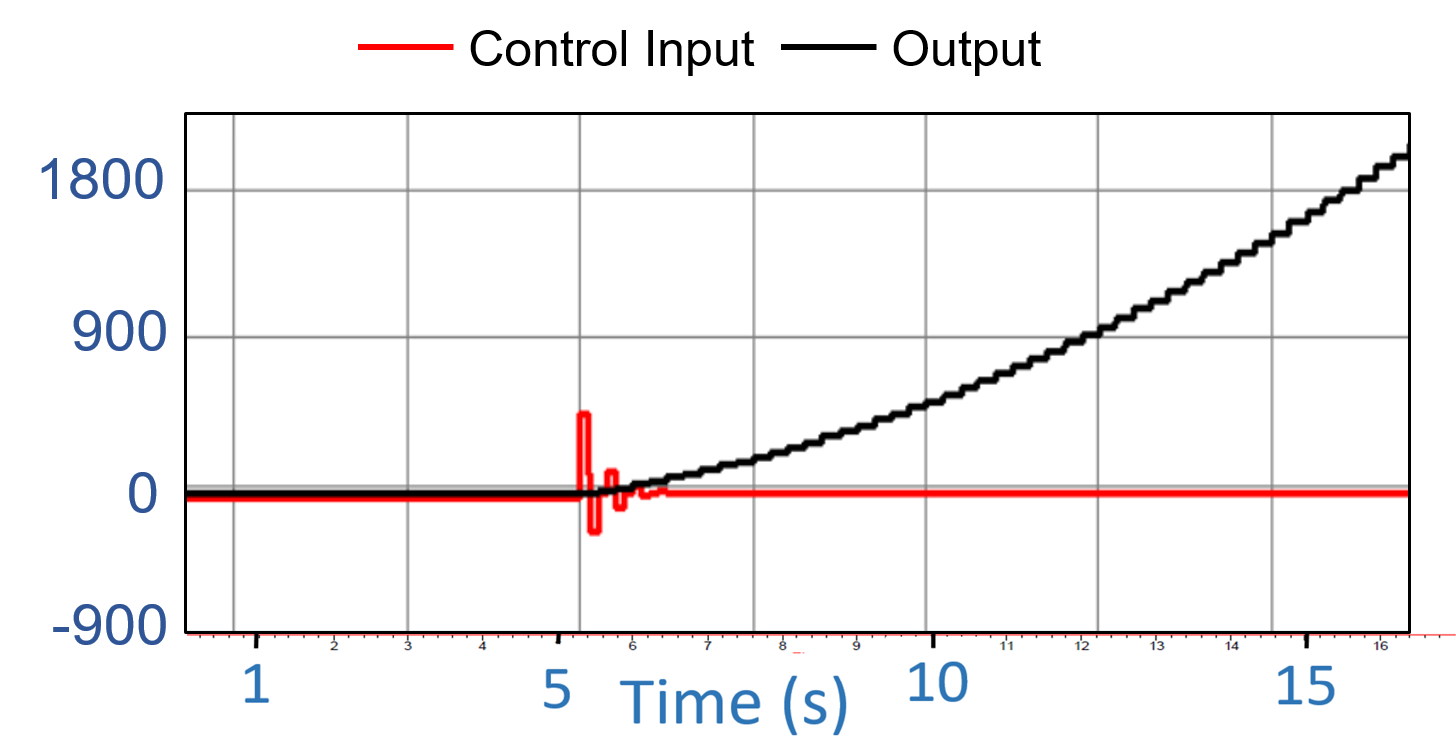}
    \caption{Bus-off attack Demonstration}
    \label{fig:busOffAttack}
    \end{subfigure}
    \begin{subfigure}[b]{0.5\columnwidth}
        \includegraphics[width=\textwidth]{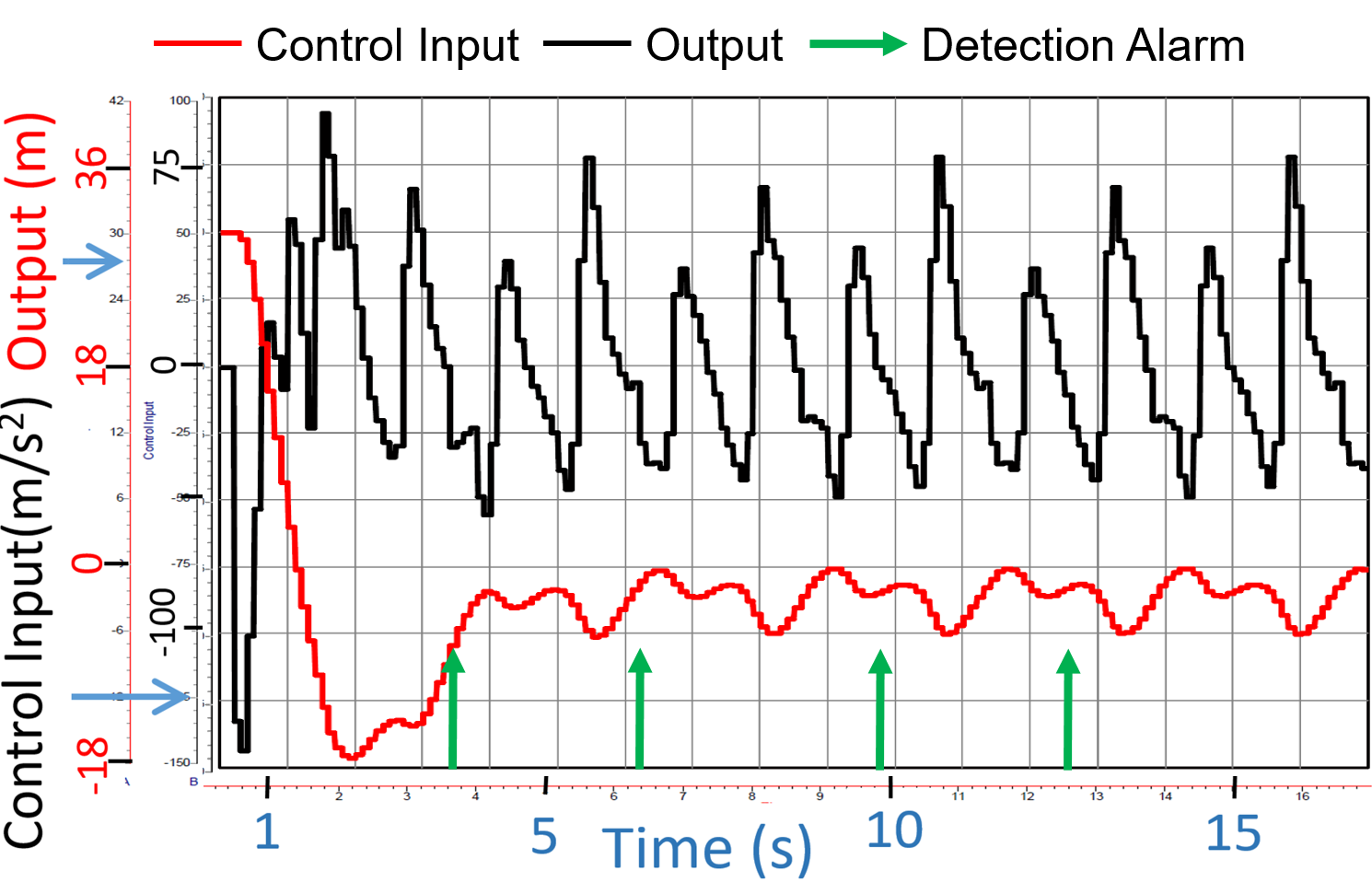}
    \caption{Detection by \hns }
    \label{fig:attackDetection}
    \end{subfigure}
 \begin{subfigure}[b]{0.5\columnwidth}
    \includegraphics[width=\textwidth]{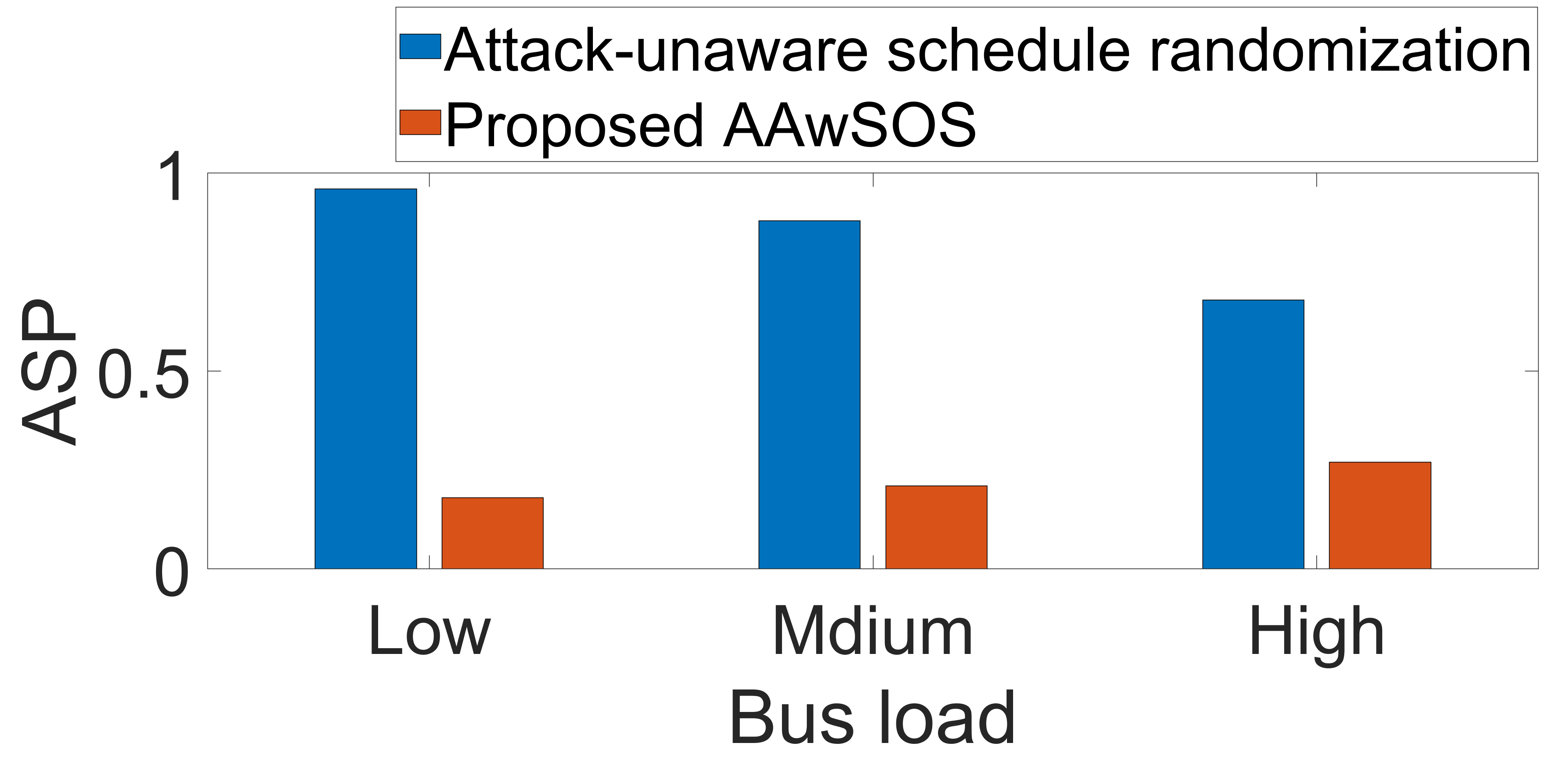}
    \caption{Comparing other strategies}
    \label{fig:comparison}
    \end{subfigure}
    \caption{Detection and Comparison.}
    \label{fig:busoffAndDetection}   
\end{wrapfigure}
In the absence of any attack-aware schedule obfuscation, a bus-off attempt on the victim message ID 0xC4 is successful (see Fig.~\ref{fig:busOffAttack}). On the other hand, the presence of \aawsos\ mandates a task execution schedule with control skips introduced at alternate instances (which respects the skip limit $1$ for TTC).

Moreover, the bus-off attack attempts are detected during the applications of $Obf_1$ (marked with the green arrow in Fig.\ref{fig:attackDetection}).

\par \noindent $\bullet$ \textit{ Comparison with Attack-unaware Schedule Randomization: } To quantify the effectiveness of our \aawsos\ compared to state-of-the-art schedule randomization techniques, we run a similar set of ASP analysis experiments (for the scenarios shown Fig.~\ref{fig:ASP Analysis}) applying typical schedule randomization techniques against a bus-off attempt. 
As discussed in Sec.~\ref{subsec:asp}, such policies are unable to reduce the total ASP ($P(AS)$) as they blindly randomize the task execution schedules. As can be seen in Fig.~\ref{fig:comparison}, such randomization policies showcase a high ASP in the presence of a low busload (the blue bar plots in Fig.~\ref{fig:comparison}). The ASP under the presence of schedule randomization reduces as the busload increase. When compared to the ASPs under proposed \aawsos\ (the orange bars in Fig.~\ref{fig:comparison}) the ASPs are significantly less. It can be observed that \aawsos suggested by our \hns algorithm showcases higher ASPs as the bus load increases. It is evident from the experiments that the modification in ECU level task schedule by \hns manages to  reduce ASPs by a higher margin when compared with  attack-unaware randomization policies like  \cite{yoon2016taskshuffler,kruger2018vulnerability}.
 

\section{Conclusion}\label{sec:conclu}
In this work we address an important security vulnerability prevalent in the automotive domain, i.e. the impact of bus-off attack on real time control tasks. We propose a lightweight algorithmic framework \hns that applies a set of attack-aware   transformations on the task execution schedules of ECUs in order to effectively obfuscate the CAN message schedule. This reduces the bus-off attack success probability and incorporates a method to detect such attempts. 
We plan to extend  this work with thorough evaluation of task scenarios executing under different real time scheduling strategies while benchmarking the attack detectability, i.e. false-alarm and true positive rates.   

\bibliographystyle{ieeetr}

\bibliography{Ref.bib}
\end{document}